\pdfoutput=1
\documentclass[11pt]{article}

\usepackage[]{emnlp2021}

\usepackage{times}
\usepackage{latexsym}

\usepackage[T1]{fontenc}
\usepackage[utf8]{inputenc}

\usepackage{microtype}

\usepackage{url}
\usepackage{ulem}
\usepackage{graphicx}
\usepackage{epstopdf}
\usepackage{amsmath}
\usepackage{color}
\usepackage[labelfont=bf]{caption}
\usepackage{placeins}
\usepackage{sidecap}
\usepackage{booktabs}
\usepackage{bm}
\usepackage{wrapfig}
\usepackage{amssymb}
\usepackage{multirow}
\usepackage{amsthm}
\usepackage{soul}
\usepackage{xcolor}
\usepackage{xspace}
\usepackage{subcaption}
\usepackage{tablefootnote}

\usepackage{mathtools}
\mathtoolsset{showonlyrefs}

\usepackage{enumitem}
\setitemize{noitemsep,topsep=0pt,parsep=0pt,partopsep=0pt}

\newcommand{\pz}{\hphantom{0}}

\newcommand{\modelmacro}[2]{#1-{\footnotesize{\textsc{#2}}}\xspace}
\newcommand{\muralbase}{\modelmacro{MURAL}{base}}
\newcommand{\murallarge}{\modelmacro{MURAL}{large}}
\newcommand{\cxcbest}{\modelmacro{DE}{T2T+I2T}}

\newcommand{\alignen}{\modelmacro{ALIGN}{base-en}}
\newcommand{\alignmling}{\modelmacro{ALIGN}{base}}
\newcommand{\alignhuge}{\modelmacro{ALIGN}{L2}}
\title{MURAL: Multimodal, Multitask Retrieval Across Languages}

\author{Aashi Jain \hspace{4pt} Mandy Guo \hspace{4pt} Krishna Srinivasan  \hspace{4pt} Ting Chen \hspace{4pt} Sneha Kudugunta \\ {\bf Chao Jia}  \hspace{4pt}  {\bf Yinfei Yang} \hspace{4pt}  {\bf Jason Baldridge} \\
        Google Research \\
        \texttt{\{aashijain, xyguo, krishnaps, iamtingchen, snehakudugunta,}\\
        \texttt{chaojia, yinfeiy, jasonbaldridge\}@google.com}
        }
\begin{document}
\maketitle
\begin{abstract}

% Prior work like ALIGN shows one can learn good image-text align model with large scale noisy image-text pairs mined from Web.
% Although the approach is easy to extend to more languages as it doesn't require any extra annotation, we found the language distribution of noisy web image-text data is highly skewed to a small group of languages.
% This leads to the model trained from such data does not perform very well for those under resourced languages.

%%%% ABSTRACT FOR FINAL VERSION %%%%
Both image-caption pairs and translation pairs provide the means to learn deep representations of and connections between languages. We use both types of pairs in MURAL (MUltimodal, MUltitask Representations Across Languages), a dual encoder that solves two tasks: 1) image-text matching and 2) translation pair matching. By incorporating billions of translation pairs, MURAL extends ALIGN \cite{jia2021scaling}--a state-of-the-art dual encoder learned from 1.8 billion noisy image-text pairs. When using the same encoders, MURAL's performance matches or exceeds ALIGN's cross-modal retrieval performance on well-resourced languages across several datasets. More importantly, it considerably improves performance on under-resourced languages, showing that text-text learning can overcome a paucity of image-caption examples for these languages. On the Wikipedia Image-Text dataset, for example, \muralbase improves zero-shot mean recall by 8.1\% on average for eight under-resourced languages and by 6.8\% on average when fine-tuning. We additionally show that MURAL's text representations cluster not only with respect to genealogical connections but also based on areal linguistics, such as the Balkan Sprachbund.

\end{abstract}

\section{Introduction}

Multilingual captions for images provide indirect but valuable associations between languages \cite{gella2017image}. \citet{burns2020eccv} exploit this to scale multimodal representations to support more languages with a smaller model than prior studies. More recent work learns cross encoder models with multitask training objectives \cite{huang2020m3p,zhou2021uc2}; in these, a single multimodal encoder attends to both inputs and exploits deep associations between images and captions. Unfortunately, such models do not support efficient retrieval \cite{geigle-etal-retrieval}, and they use object detection, machine translation, bilingual dictionaries and many losses. In contrast, multimodal dual encoders can be learned directly on noisy, massive image-caption datasets using a simple loss based on in-batch bidirectional retrieval \cite{jia2021scaling,radford2021learning}. These support efficient retrieval via approximate nearest neighbors search \cite{avq_2020} and can predict similarity within and across modalities \cite{parekh-etal-2021-crisscrossed}.

%M3P learns using several pretraining tasks, including Masked Language and Masked Region modeling--both of which take combined image and text input. UC2 is another cross encoder, but it uses Masked Region-to-Token Modeling and Visual Translation Language Modeling, and it relies on multilingual image-text pairs obtained via machine translation. 

%The effectiveness of neural language models grows with scale of both model and data \cite{kaplan2020scaling}, but for many under-resourced languages there simply is very little raw text available, compared to well-resourced languages like English. Additional modalities such as images provide valuable associations between languages and language-neutral artifacts \cite{gella2017image}, so adapting a multilingual text encoder with multilingual image-text data may improve representations derived for under-resourced languages, beyond the value of learning from just their language portions.

\begin{figure}
    \begin{center}
        \includegraphics[trim=0 0 0 0,clip,scale=0.5]{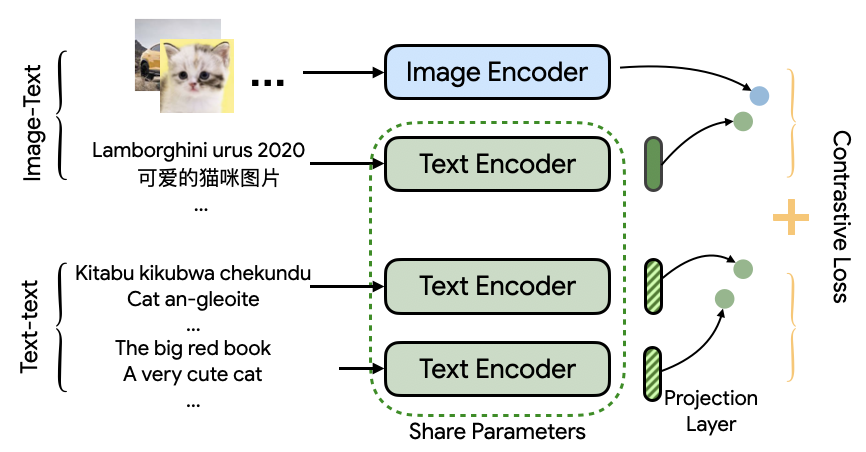}
        \caption{MURAL learns encoders for both language and images by combining both image-text matching and text-text matching tasks, using scalable dual encoder models trained with contrastive losses.}
    \label{fig:teaser}
    \vspace{-0.2in}
    \end{center}
    
\end{figure}

\begin{table*}[t]
%\small
\centering
\scalebox{1.0}{
	\begin{tabular}{lrrrrrrr}
	%\toprule
	Name &  Train-I & Train-T & Dev-I & Dev-T & Test-I & Test-T & \#Langs\\
	\midrule
    EOBT Pairs  & - & 500m & - & - & - & - & 124 \\
	MBT Pairs$^\dagger$ & - & 6b & - & - & - & - & 109 \\
	\hline
    CC12m & 12m & 12m & - & - & - & - & 1 \\ 
	Alt-Text$^\dagger$ & 1.8b & 1.8b & - & - & - & - & 110 \\
	
	XTD & - & - & - & - & 1k & 1k & 7\\
    Multi30k & 29k & 145k & 1k & 5k & 1k & 5k & 4 \\
    MS-COCO & 82k & 410k & 5k & 25k & 5k & 25k & 1\\
	STAIR & 82k & 410k & 5k & 25k & 5k & 25k & 1\\
	WIT & 11.4m & 16m & 5/3/1k & 5/3/1k & 5/3/1k & 5/3/1k & 108\\
	%\bottomrule                    
	\end{tabular}
}
\caption{Dataset statistics. Counts are per language, except that Alt-Text and WIT training counts aggregate over all languages. WIT text counts are for reference descriptions.  ({\it Key}: I=Image, T=Text; $^\dagger$: indicates internal datasets); see Section \ref{sec:data} for abbreviations and further details on each dataset.)} 
\label{table:data_stats}
\end{table*}

With \textbf{MURAL}: MUltimodal, MUltitask Representations Across Languages (Fig. \ref{fig:teaser}), we explore dual encoder learning from both image-caption and translation pairs at massive scale: 6 billion translation pairs \cite{feng2020language} and 1.8 billion image-caption pairs \cite{jia2021scaling}. We particularly seek to improve performance for under-resourced languages. Addressing this was infeasible until now because existing multilingual image-text datasets---Multi30k \cite{elliott2016multi30k}), STAIR \cite{yoshikawa2017stair}, and XTD \cite{aggarwal2020towards}--support only high-resource languages.  However, the recent Wikipedia Image-Text (WIT) dataset \cite{srinivasan2021wit}, which covers 108 languages, addresses this gap.

Our results, as a whole, demonstrate that ALIGN, a state-of-the-art multimodal dual encoder, is improved by adding a bitext ranking objective~\cite{yang-ijcai2019-746} (=MURAL). The latter matches zero-shot image-text retrieval performance on well-resourced languages, and it dramatically improves performance on under-resourced languages. For XTD, MURAL improves recall@10 by 4\% on average. On WIT zero-shot, MURAL improves mean recall by 1.7\% on average for nine well-resourced languages, and by 8.1\% for eight under-resourced ones. After fine-tuning on WIT, MURAL mean recall is 1.8\% and 6.8\% better than ALIGN, on average, for well-resourced and under-resourced languages, respectively.

We also show that the resulting dual encoder model can outperform more complex cross-encoder baseline models by a wide margin, thus obtaining stronger performance from models that support scalable retrieval. Our largest model, \murallarge, improves mean recall for zero-shot retrieval by 47.7\% on average for four languages in Multi30k over M3P \cite{huang2020m3p}. It improves mean recall by 5.9\% over UC2 \cite{zhou2021uc2} for the fine-tuning setting of Multi30k.  \murallarge also improves over a strong \textit{translate-test} baseline on WIT in a zero-shot setting for well-resourced languages by 13.2\% and for under-resourced ones by 9.6\%.

We report results on Crisscrossed Captions (CxC) \cite{parekh-etal-2021-crisscrossed}, which additionally provides image-text, text-text, and image-image similarity ratings. \murallarge obtains the highest scores to date on CxC text{$\rightarrow$}text and image{$\rightarrow$}image retrieval. Our small ALIGN model and \murallarge model tie for best Semantic Image Similarity, which measures the correlation between model rankings and human rankings over image-image pairs.

Finally, we show that multilingual representations learned in MURAL form clusters which are influenced from areal linguistics and contact linguistics, in addition to previously shown genealogical relationships \cite{kudugunta2019investigating}.
%\input{020_related} % Related work section is optional. Ideally we should embed related work in intro.
% \begin{figure*}
%     \centering
%     % \subfloat{{\includegraphics[width=7cm]{images/WIT_Data_Distribution.png} }}
%     % \qquad
%     \subfloat{{\includegraphics[width=.9\textwidth]{images/alt_log_linear.png} }}
%     \caption{Data distribution across languages for Alt-Text. The enlarged plot is at log-scale to provide a better view of under-represented languages. The embedded chart is at a normal scale to better convey the skew toward the well-resourced languages.}
%     \label{fig:wit_alttxt_stats}
% \end{figure*}

\begin{figure*}[t]
    \centering
    \begin{subfigure}[b]{0.49\textwidth}
         \centering
         \includegraphics[width=\textwidth]{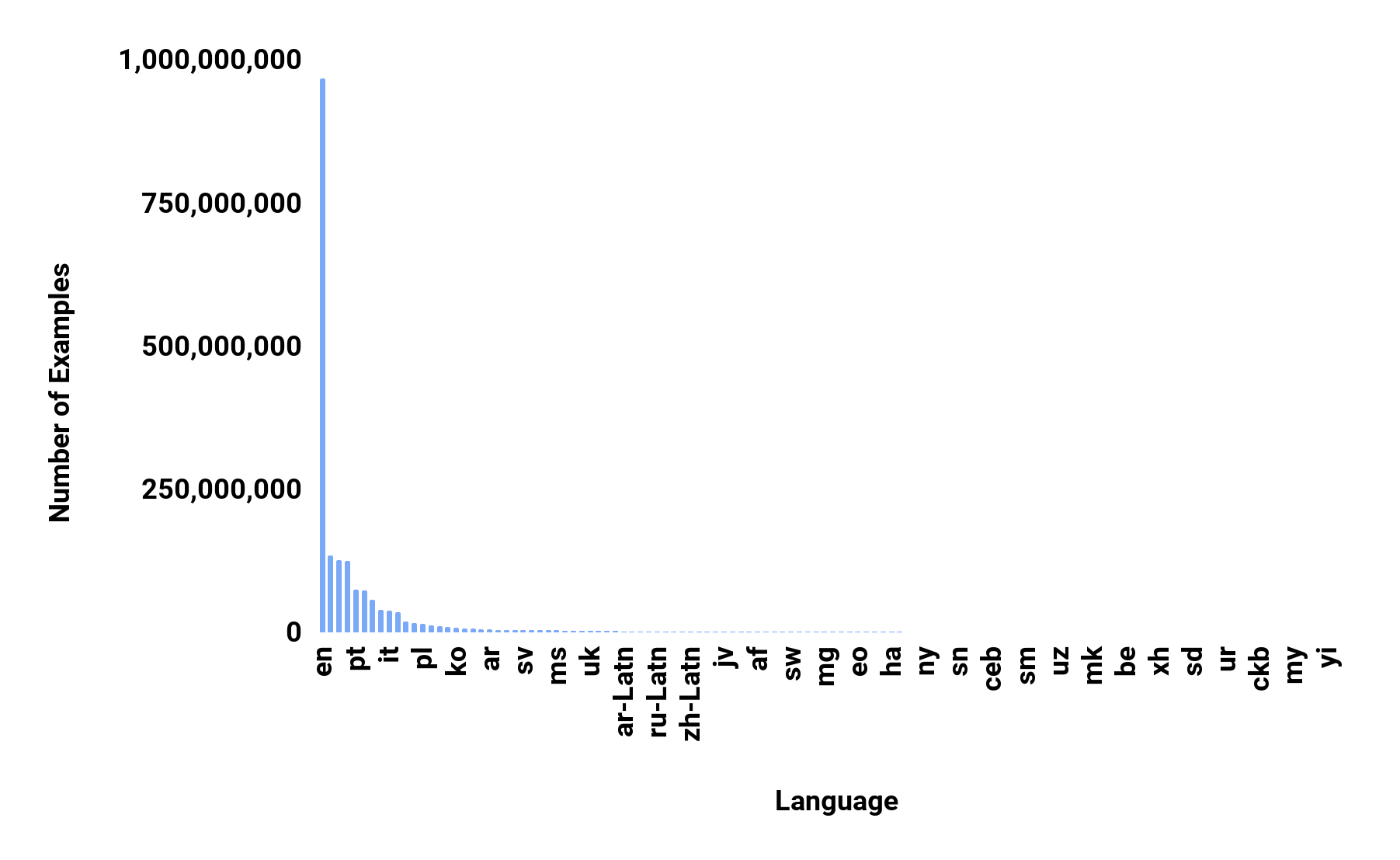}
         \label{fig:y equals x}
     \end{subfigure}
    % \hfill
    \begin{subfigure}[b]{0.49\textwidth}
         \centering
         \includegraphics[width=\textwidth]{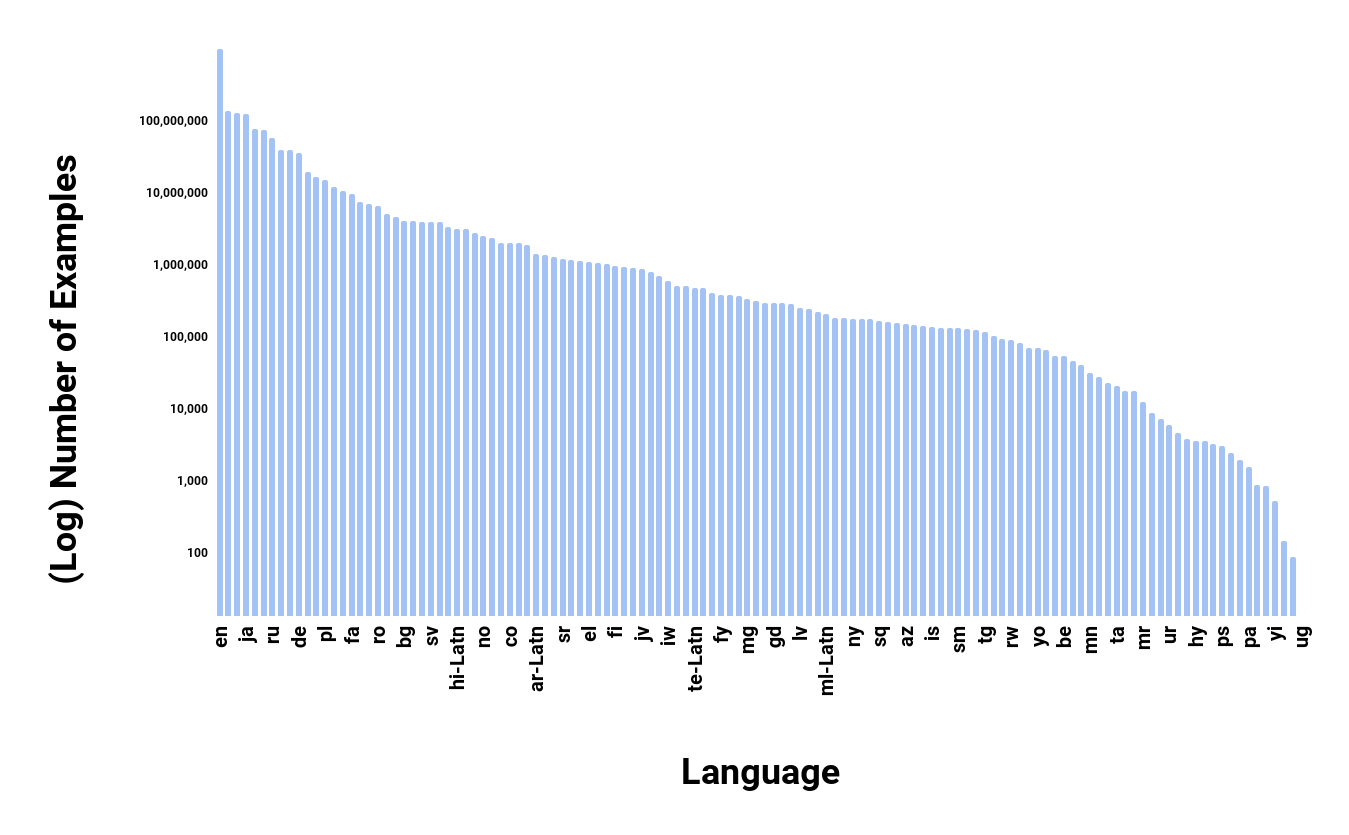}
         \label{fig:y equals x}
    \end{subfigure}
     
    \caption{Alt-Text language distribution: (left) linear scale, which clearly conveys the skew toward well-resourced languages; (right) log-scale, which provides a better view of under-represented languages.}
    \label{fig:wit_alttxt_stats}
\end{figure*}

% \krishnaps{Since we are using the WIT-train set, here are the numbers:
% Total examples: 37,139,639
% Unique images: 11,414,134 (we use this)
% attr desc: 34,827,740		
% ref desc: 16,977,508 (we use this)
% alt text caption: 5,340,623}
% We should explain that WIT's text here means the reference description.

\section{Data}
\label{sec:data}

For training, we use both publicly available datasets and internal ones that are much larger. We evaluate on many publicly available image captioning datasets. Table \ref{table:data_stats} summarizes their statistics.

\subsection{Training datasets}

\textbf{Conceptual 12M} (CC12M) \citet{changpinyo2021conceptual} is a publicly available image captioning dataset in English with 12 million pairs obtained from web images and their corresponding alt-text descriptions. CC12M loosens the strong quality filters on the earlier Conceptual Captions (CC3M) dataset \cite{sharma-etal-2018-conceptual} to obtain greater scale.

The multilingual version of \textbf{Alt-Text} \cite{jia2021scaling} is a noisy dataset with 1.8 billion images and their alt-text descriptions, covering 110 languages. Alt-Text is minimal filtered; this increases the scale and diversity of both images and languages. Fig. \ref{fig:wit_alttxt_stats}, which gives the distribution over all languages: over half the captions are English, and the top fifth of languages covers 95\% of captions, so many languages still have \textit{relatively} fewer examples.

We create an \textbf{Ensemble of Open Bilingual Translation (EOBT) Pairs} dataset by combining publicly available datasets, including Europarl \cite{koehn2005europarl}, Paracrawl \cite{espla-etal-2019-paracrawl}, Wikimatrix \cite{schwenk2019wikimatrix}, and JW300 \cite{agic2020jw300}---see Appendix \ref{appendix:eobt} for a full list. EOBT has $\approx$500 million pairs across all languages.

\citet{feng2020language} mine translations from the web; we call their dataset as \textbf{Mined Bilingual Translation (MBT) Pairs}. It has 6 billion pairs (up to 100 million per language) for 109 languages.

\subsection{Evaluation datasets}

\textbf{Flickr30K} \cite{young2014image} has 31k images, with five English captions per image. \textbf{Multi30K} extends Flickr30k with German, French, and Czech captions. \citet{elliott2016multi30k} introduces German annotations by 1) translating some Flickr30k English captions and 2) crowdsourcing new German captions for Flickr30K images. Following prior work \cite{burns2020eccv}, we report results on the independent 5 captions/image split. \citet{elliott2017findings} and \citet{barrault2018findings} further extend the dataset by collecting human translations of English Flickr30k captions to French and Czech. 

\textbf{MS-COCO} \cite{lin2014microsoft} also has five human generated English captions per image. We report results on both the 1k and 5k splits defined by \citet{karpathy2015deep}.  The \textbf{STAIR} dataset \cite{yoshikawa2017stair} adds human crowd-sourced Japaneses captions for MSCOCO images.

% \vspace{5pt} \noindent \textbf{XTD10} dataset is a 1k image-text test-only extension of MSCOCO dataset in 7 different languages (Korean, Russian, Polish, Spanish, Turkish, Chinese, Japanese) introduced by \citet{aggarwal2020towards}. It is important to note that the split used by the authors is different from the widely used split for MSCOCO.

\textbf{XTD} \citet{aggarwal2020towards} created the Cross-lingual Test Dataset for evaluating multimodal retrieval models. XTD does not include any training examples, but it supports retrieval evaluation on seven diverse languages.

The large-scale \textbf{Wikipedia Image Text (WIT)} dataset \cite{srinivasan2021wit} is mined from Wikipedia, covering 108 languages. The validation and test splits for WIT are not publicly available, so we partition the training data to construct new splits for WIT.\footnote{\url{https://github.com/google-research-datasets/wit}} For most languages, we use 5k image-text pairs each for validation and test, but for less well-resourced languages, we use 3k or 1k pairs. See Appendix \ref{appendix:wit} for details.

\textbf{Crisscrossed Captions (CxC)} \cite{parekh-etal-2021-crisscrossed} extends the English MSCOCO 5k dev and test sets with human similarity annotations for both intra- and inter- modal tasks. As such, CxC supports evaluation for both inter-modal (image-text) and intra-modal (text-text, image-image) retrieval tasks, and correlation measures that compare model rankings with rankings derived from human similarity judgments (again, for image-text, image-image and text-text comparisons).

% \footnote{We will release these splits to facilitate reproducibility.}
% NOTE(krishnaps): I commented the above footnote - We might not want to commit to this. The last time I had a discussion with the team, everyone felt it would be confusing to have another set of test/val in addition to the competition.

% TODO(krishnaps):
%\todo{These are placeholder images with the totals from public data. This will be replaced with the more accurate count of how much we exactly ended up using.}
% scale=0.1
% width=\textwidth
%\begin{figure}
%    \begin{center}
%\includegraphics[scale=0.13]{emnlp2020-templates/images/WIT_Data_Distribution.png}
%        \caption{WIT Dataset Image-Text Distribution Per Language.}
%    \label{fig:wit_stats}
%    \vspace{-0.2in}
%    \end{center}
%\end{figure}
%
%\begin{figure}
%    \begin{center}
%        \includegraphics[scale=0.11]{emnlp2020-templates/images/AltText_Data_Distribution.png}
%        \caption{Alt-Text Dataset Distribution Per Language.}
%    \label{fig:alttext_stats}
%    \vspace{-0.2in}
%    \end{center}
%\end{figure}

\section{Models}

ALIGN \cite{jia2021scaling} is a family of multimodal dual encoders that learn to represent images and text in a shared embedding space. ALIGN's encoders are trained \textit{from scratch} on image-text pairs via an in-batch normalized softmax loss (contrastive learning). This loss encourages the model to encode positive image-text pairs closer to each other while pushing away in-batch negative pairs.

ALIGN delivers state-of-the-art results for several datasets; however, the \textit{Alt-Text} data used to train it is heavily skewed towards well-resourced languages (see Fig. \ref{fig:wit_alttxt_stats}). This imbalance reduces ALIGN's ability to represent under-resourced languages; we address that here by using more representative text-text translation pairs mined at scale from the web.

%Beyond image-text pairs, recent progress has been made in reducing the gap among these languages in text-only datasets like Machine Translation datasets, Monolingual text datasets. This motivates us to use the text-text datasets mined from the Web, which could potentially help the model learn better representations for those languages for which image-text pairs are not yet available. 

%\subsection{Multimodal Multitask Contrastive Learning}
\subsection{MURAL}
\label{sec:mural}

MURAL extends ALIGN with a \textit{multitask} contrastive learning objective that adds text-text contrastive losses to the image-text ones. MURAL is trained simultaneously with two tasks of image-text (i2t) matching and text-text (t2t) matching. The text encoder is shared between these two tasks to allow transfer of multilingual learning from the text-text task to cross-modal representations. The resulting loss function is the sum of losses from both tasks.

\textbf{Weighting of i2t and t2t tasks} in the loss function \cite{parekh-etal-2021-crisscrossed} allows the tasks to be balanced. We experiment with different weights for both tasks; our main focus is cross-modal retrieval, so we weigh the image-text task higher than the text-text task. We use the following loss function: 
\begin{equation}
\mathcal{L} = w_{i2t} * (\mathcal{L}_{i2t} + \mathcal{L}_{t2i}) + w_{t2t} *  (\mathcal{L}_{r2l} + \mathcal{L}_{l2r})
\end{equation}

\noindent
Each loss is an in-batch softmax of the form:

\begin{equation}
%\mathcal{L}_{i2t} = -\frac{1}{N}\sum^{N}_{i} \log \frac{\exp(x^T_i y_i/\tau)}{\sum^{N}_{j=1}\exp(y^T_i x_j/\tau)}
\mathcal{L}_{i2t} = -\frac{1}{N}\sum^{N}_{i} \log \frac{\exp(\mathrm{sim}(\bm x_i, \bm y_i)/\tau)}{\sum^{N}_{j=1}\exp(\mathrm{sim}(\bm x_i, \bm y_j)/\tau)}
\end{equation}

\noindent
%where $\bm x_i$ and $\bm y_j$ are normalized embeddings of the image of the $i$-th pair and the text in the $j$-th pair, respectively. 
where $\bm x_i$ and $\bm y_j$ are embeddings of the image in the $i$-th pair and the text in the $j$-th pair, respectively. $\mathrm{sim}(\bm x,\bm y) = \bm x^\top \bm y / \lVert\bm x\rVert \lVert\bm y\rVert$ denotes the dot product between $\ell_2$ normalized $\bm x$ and $\bm y$ (cosine similarity).
$N$ is the batch size. $\tau$ is the temperature to scale the logits. We use a similar construction for $\mathcal{L}_{t2i}, \mathcal{L}_{r2l}$, and $\mathcal{L}_{l2r}$, where $l$ is left-text and $r$ is right-text. The softmax temperature is shared between $\mathcal{L}_{i2t}$ and $\mathcal{L}_{t2i}$, and is learned with initial value 1.0. In $\mathcal{L}_{r2l}$ and $\mathcal{L}_{l2r}$, the temperature is fixed to 0.01. Following \citet{feng2020language}, we use additive margin 0.3 in $\mathcal{L}_{r2l}$ and $\mathcal{L}_{l2r}$. 

\textbf{Task-specific projection heads} that transform encoder representations before computing cosine similarity between inputs can improve contrastive learning~\cite{chen2020simple}. Similar designs have also been used for a traditional multitask setting \cite{guo2019autosem}. In MURAL, we use two single-layer, task-specific projection heads above the text encoder: one transforms the text embedding for image-text contrastive loss, and the other for text-text contrastive loss (more details in \ref{appendix:modeling}).

% We experiment with different layers of projection heads, e.g. 1 Fully Connected (FC) layer and a Multi-Layer Perceptron with non-linearity in between the FC layers. Empirically, we find that MURAL learns better image-text representations when using single layer projection heads on top of the text-encoder, one per task.

% \vspace{5pt} \noindent \textbf{Similarity with Soft-constraint via LaBSE}
% Training with translation pairs dataset can be expensive in terms of both time, space, and compute resources training. With better universal sentences representation publicly avaialbe, we show that these can be leveraged to train multi-modal representations that help in boosting the performance on under-resourced languages. This is a trade-off that can be explored by those who are willing to sacrifice representation quality for computational resources.

\textbf{Fine-tuning: single-task vs. multi-task.} Our primary goal with MURAL is to improve zero-shot performance by learning with both image-text \textit{and} text-text pairs. Nevertheless, fine-tuning has a large impact on performance for any given dataset. After initial experiments, we find that single-task fine-tuning using image-text pairs performed slightly better than multitask finetuning using co-captions. For further discussion on this comparison, see Appendix \ref{appendix:modeling}. For all models, we report results using single-task fine-tuning using any available training image-text pairs for a given dataset.
\begin{table*}[t]
\centering
\scalebox{.75}{
      \begin{tabular}{llllc@{\hskip 0.3in}cccc@{\hskip 0.2in}cc@{\hskip 0.2in}cc}
            %\toprule
            & &  & &  & \multicolumn{4}{c}{\textbf{Multi30K}} & \multicolumn{2}{c}{\textbf{MSCOCO 1K}} & \multicolumn{2}{c}{\textbf{MSCOCO 5K}}\\
            \cmidrule(l){6-9} 
            \cmidrule(l){10-13}
            & & \bf Model & \bf Data & \bf Type & en   & de   & fr   & cs   & en & ja & en & ja \\
            \midrule
            \multirow{7}{*}{\rotatebox[origin=c]{90}{Zero-shot}} % rotate is optional if we need more space
            & (1) & M3P & CC3m+Wiki & CE & 57.9 & 36.8 & 27.1 & 20.4 & 63.1 & 33.3 & - & -  \\
            
            & (2) & \alignmling &  TrTrain(AT-en) & DE & 82.0 & 75.2 & 74.7 & 68.2 & 77.1 & 70.6 & 55.9 & 46 \\
             
            & (3) & \alignen & AT-en$\rightarrow$translate-test & DE & 84.3 & 78.9 & 78.3 & 71.1 & 80.0 & 71.5 & 60.6 & 51.9 \\

            & (4) & \alignmling & AT & DE & 83.3 & 75.0 & 74.2 & 47.9 & 79.5 & 70.9 & 59.6 & 53.9\\
            %& (5) & \muralbase & CC12m+EOBT & DE & 76.2 & \textcolor{red}{29.4} & 44.4 & \textcolor{red}{28.3} & 75.4 & \textcolor{red}{21.7} & 53.7 & \textcolor{red}{\pz9.9} \\
            & (5) & \muralbase & TrTrain(CC12m)+EOBT & DE & 80.9 & 76.0 & 75.7 & 68.2 & 78.1 & 72.5 & 58.0 & 49.7\\ % https://mldash.corp.google.com/experiments/5885587308363892925#scalars This model is trained 600k steps as we have more data with translated versions.
            & (6) & \muralbase & AT+MBT & DE  & 82.4 & 76.2 & 75.0 & 64.6 & 79.2 & 73.4 & 59.5 & 54.4\\
            & (7) & \murallarge & AT+MBT & DE & 89.2 & \bf 83.5 & \bf 83.1 & \bf 77.0 & \bf 84.4 & \bf 81.3 & 67.7 & \bf 64.6 \\
            
            % Soft-constraint (ALIGN-sm) & 78.4 & \multicolumn{1}{c}{71.1} & \multicolumn{1}{c}{72} & \multicolumn{1}{c}{52.8} & 54.1   & 39  \\
            & (8) & \alignhuge & AT-en & DE & \bf 92.2 & - & - & - & - & - & \bf 70.9 & - \\ 
            \midrule
            \multirow{8}{*}{\rotatebox[origin=c]{90}{Fine-tuned}}
            %& EmbN & & DE & 72.0 & 60.3 & 54.8 & 46.3 & 76.8* & 73.5* & - & -  \\
            %& PAR.EmbN & & DE & 69.0 & 62.6 & 60.6 & 54.1 & 78.3* & 74.8* & - & -  \\
            %& S-LIWE & & DE & 76.3 & 72.1 & 63.4 & 59.4 & 80.9* & 70.0* & - & - \\
            % & MULE & & DE & 70.3 & 64.1 & 62.3 & 57.7 & -- & -- & &  \\
            & (9) & SMALR & \textit{no pretraining} & DE & 74.5 & 69.8 & 65.9 & 64.8 & \pz81.5$^\dagger$ & \pz77.5$^\dagger$ & - & -  \\
            & (10) & M3P & CC3m+Wiki &  CE & 87.7 & 82.7 & 73.9 & 72.2 & \pz88.7$^\dagger$ & \pz87.9$^\dagger$ & - & -  \\
            & (11) & UC2 & TrTrain(CC3m) & CE  & 88.2 & 84.5 & 83.9 & 81.2 & \pz88.1$^\dagger$ & \pz87.5$^\dagger$ & - & -   \\
            & (12) & \alignmling & TrTrain(AT-en) & DE & 92.2 & 88.5 & 88.1 & 84.5 & 89.0 & 87.5 & 74.8 & 72.5 \\
            & (13) & \alignmling & AT & DE & 92.3 & 88.3 & 78.8 & 81.4 & 89.2 & 86.7 & 76.1 & 74.1 \\
            % & MURAL & DE & 90.3 & 85.6 & {85.2} & {81.5} & & & 78.6 & 60.9  \\
            %& (5) & \muralbase & CC12m+EOBT & DE & 87.0 & 82.0 & 78.5 & 71.1 & 84.6 & 71.8 & 66.2 & 50.4 \\
            & (14) & \muralbase & TrTrain(CC12m)+EOBT & DE & 91.0 & 87.3 & 86.4 & 82.4 & 89.4 & 87.4 & 73.7 & 71.9 \\
            & (15) & \muralbase & AT+MBT & DE & 92.2 & 88.6 & 87.6 & 84.2 & 88.6 & 88.4 & 75.4 & 74.9 \\
            & (16) & \murallarge & AT+MBT & DE & 93.8 & \bf 90.4 & \bf 89.9 & \bf 87.1 & \bf 92.3 & \bf 91.6 & 81.2 & \bf 81.3 \\
            & (17) & \alignhuge & AT-en & DE & \bf 96.0 & - & - & - & - & - & \bf 83.4 & - \\ 
            %\bottomrule
      \end{tabular}
      }
      \caption{Mean recall on standard datasets.
      $^\dagger$: Numbers from UC2 paper; these were fine-tuned on MSCOCO-CN \cite{Li2019COCOCNFC}, which has a different split than {\it en} and {\it ja}, resulting in possible train/test infiltration. SMALR MSCOCO 1K results use a different test split. {\scriptsize ({\it Key}: AT=Alt-Text dataset, DE=Dual Encoder, CE=Cross Encoder, TrTrain=translate-train)}}
\label{tab:results-standard-datasets}
\end{table*}

\subsection{Model variants}

\citet{jia2021scaling} trains a very large model, \textbf{\alignhuge}, that uses EfficientNet-L2 \cite{Tan2019EfficientNetRM} as image encoder and BERT-Large \cite{devlin-etal-2019-bert} as the text encoder. It was trained on English-only Alt-Text data. We explore smaller models and fewer training epochs to study various strategies more efficiently. For this, we use directly comparable \textbf{\alignmling} and \textbf{\muralbase} models: both use EfficientNet-B5 for image encoding and BERT-Base for text. \muralbase also uses text-text learning and an additional projection head for the image-text task (see Sect. \ref{sec:mural}). We also consider \textbf{\murallarge}, which uses Efficient-B7 and BERT-Large. \alignmling and \muralbase have $\approx$300M parameters,  \murallarge has $\approx$430M, and \alignhuge has $\approx$840M parameters. Appendix \ref{appendix:modeling} gives more details.

Following ALIGN \cite{jia2021scaling}, we use LAMB optimizer \cite{You2020LargeBO} with a weight decay ratio of 1e-5. For \alignmling and \muralbase, we train our models on 128 Cloud TPU V3 cores with a global batch size of 4096. The image-text task uses a learning rate of 1e-3 and the text-text task uses 1e-4. Both learning rates are linearly warmed up from zero to their final values in 10k steps and then decayed linearly to zero in 600k steps. This corresponds to only around 1.4 epochs of the Alt-Text dataset and 0.4 epochs of the MBT dataset. \murallarge is trained on 512 TPU cores (4x larger samples used in training).

We build a 250k word-piece vocabulary from the Alt-Text training data,\footnote{The vocabulary is built using the standard wpm library from \href{https://github.com/tensorflow/text/blob/master/tensorflow_text/tools/wordpiece_vocab/generate_vocab.py}{tensorflow\_text}.} which is kept the same in all our experiments to control the changing factors.

\subsection{Baseline Strategies}

Our main goal is to explore the potential of large, diverse translations pairs for learning better multimodal encoders, including a \textit{single} multilingual text encoder. We compare this strategy to the well-established, effective baselines of \textbf{translate-train} and \textbf{translate-test} using a strong Neural Machine Translation (NMT) system\footnote{\url{https://cloud.google.com/translate}} \cite{yang-etal-2019-paws}. 

\textbf{Translate-train:} To reduce the heavy bias toward English and to support other languages for models training only on image-text pairs (e.g. for ALIGN), we artificially create image-text pairs by using the NMT system to translate English texts to other languages.\footnote{Refer to appendix \ref{appendix:translate_train} for more details.} These additional pairs are then used to train the model -- a core strategy used in UC2 \cite{zhou2021uc2}.

\textbf{Translate-test}: An alternative strategy is to train a high-performing English model and then translate non-English inputs into English, which are then encoded for cross-modal retrieval at test time.

Both strategies are highly dependent on the quality of NMT system, the languages it supports, while also incurring additional cost and complexity \footnote{Translating a text query with 10 tokens adds additional latency of upto 400ms in run on CPU with a batch size of 1, }.

%With advances in automatic translation quality, \textbf{augmentation using Neural Machine Translation (NMT)} is now a a viable strategy. With \textit{translate-train}, image-text datasets are expanded by translating captions into other languages. This can boost performance, but it is an expensive operation that is highly dependent on the quality of machine translation system used. Conversely, one can train a monolingual model and translate examples at test time (\textit{translate-test}), which is also dependent on NMT output quality.
        
        \section{Results}
        
        %We first compare MURAL to previous work on existing, popular datasets that contain training and/or evaluation examples that emphasize high-resource languages. We additionally evaluate MURAL on the challenging new Wikipedia Image-Text dataset, using new splits of the data to standardize evaluation across languages and accommodate the differing amounts of examples available for well-resourced languages to under-represented ones.
        
        We focus on:
        
        \begin{enumerate}
            \item Evaluating the impact of MURAL's text-text loss by comparing \alignmling and \muralbase, \textbf{especially for under-resourced languages}.
            \item Understanding the impact of training data scale by comparing Alt-Text+MBT to CC12M+EOBT.
            \item Situating our best model, \murallarge, with respect to previous work.
        \end{enumerate}
        \noindent
        We number the rows in our results tables to ease reference in our discussion and across tables.
        
        \textbf{Multi30k and MSCOCO.} Table \ref{tab:results-standard-datasets} compares MURAL and previous results \cite{burns2020eccv, huang2020m3p,zhou2021uc2, jia2021scaling} in both zero-shot and fine-tuned settings.
        
        The additional text-text task used by \muralbase improves zero-shot performance on Czech, a relatively lower-resourced language, by a large margin over \alignmling (4 vs 6), 47.9 $\rightarrow$ 64.6,  while nearly matching or somewhat exceeding performance on higher-resource languages. 
        
        Large, noisy pre-training greatly reduces the need for fine-tuning. M3P sees huge performance gains by fine-tuning\footnote{Fine-tuned on Multi30k and MSCOCO combined, trained for 40k steps and learning rate sweeping of 1e-5, 5e-5, and 1e-4. Other hyperparameters are kept the same.} (1 vs 10), sometimes 3x the zero-shot performance. Both \alignmling and \muralbase see large gains, but their zero-shot performance is already near M3P's fine-tuned performance for highly resourced languages. \murallarge's zero-shot (7) actually exceeds M3P's fine-tuned performance (10) and almost matches UC2's fine-tuned performance (11).

        \begin{table*}[t]
        \scalebox{0.68}{
            \begin{tabular}{lll|ccccccccc|ccccccccc}
                & & & \multicolumn{9}{c}{Well-resourced} & \multicolumn{8}{c}{Under-resourced}\\
                & & Model & en & de   & fr  & cs & ja & zh & ru & pl & tr & tg    & uz  & ga  & be  & mg  & ceb & ht  & war \\
                \hline
                \multirow{4}{*}{\rotatebox[origin=c]{90}{Zero-shot}}
                & (3) & \alignen & 46.5 & 33.9 & 42.3 & 32.4 & 29.9 & 36.2 & 40.1 & 39.2 & 40.5 & 30.0 & 23.4 & 26.1 & 27.3 & 33.6 & 34.9 & 41.6 & n/a$^*$ \\
                & (4) & \alignmling & 46.7 & 33.5 & 45.0 & 26.5 & 33.6 & 35.2 & 30.9 & 29.9 & 31.4 & 21.2 & 15.6 & 12.9 & 8.9 & 23.9 & 31.0 & 33.1 & 24.0  \\
                & (6) & \muralbase & 46.4 & 33.9 & 44.8 & 31.5 & 34.3 & 35.6 & 33.7 & 33.2 & 34.7 & 35.3 & 24.1 & 20.8 & 21.4 & 33.0 & 35.7 & 39.1 & 26.1\\
                % & (7) & \murallarge & \bf{60.2} & \bf{46.0} & \bf{58.8} & \bf{44.1} & \bf{37.0}$^\dagger$ & 28.8$^\dagger$ & \bf{46.5} & \bf{46.0} & \bf{49.2} & \bf{42.3} & \bf{34.0} & \bf{30.5} & \bf{33.0} & \bf{46.0} & \bf{45.1} & \bf{52.8} & \bf{37.3} \\
                & (7) & \murallarge & \bf 60.7 & \bf 46.1 & \bf 60.0 & \bf 43.6 & \bf 48.1 & \bf 49.9 & \bf 45.7 & \bf 45.8 & \bf 49.8 & \bf 45.7 & \bf 33.7 & \bf 30.8 & \bf 33.4 & \bf 45.6 & \bf 45.6 & \bf 52.4 & \bf 37.7\\
                \hline
                \multirow{4}{*}{\rotatebox[origin=c]{90}{Fine-tuned}}
                & (21) & \alignen &66.4 & 48.8 & 58.5 & 44.7 & 40.2 & 48.2 & 55.2 & 52.0 & 58.0 & 47.0 & 29.6 & 32.7 & 37.7 & 44.2 & 48.4 & 53.5 & n/a* \\
               
                & (18) & \alignmling & 75.6 & 69.2 & 76.2 & 65.5 & 64.4 & 78.2 & 68.3 & 68.3 & 75.0 & 53.0 & 36.3 & 35.8 & 50.3 & 45.0 & 72.4 & 62.5 & 78.1 \\
                & (19) & \muralbase & 77.1 & 70.0 & 77.2 & 68.4 & 64.8 & 79.6 & 70.8 & 70.7 & 78.2 & 64.2 & 44.1 & 41.9 & 59.3 & 55.1 & 76.4 & 67.6 & 79.0 \\
                & (20) & \murallarge & \bf 82.4 & \bf 76.3 & \bf 83.3 & \bf 74.5 & \bf 71.9 & \bf 86.7 & \bf 77.4 & \bf 77.4 & \bf 85.7 & \bf 72.9 & \bf 53.5 & \bf 51.4 & \bf 69.8 & \bf 62.3 & \bf 82.3 & \bf 76.7 & \bf 84.2\\
              \end{tabular}
              }
              \caption{Mean Recall on WIT for English (en); German (de); French (fr); Czech (cs); Japanese (ja); Chinese (zh); Russian (ru); Polish (pl); Turkish (tr); Tajik (tg); Uzbek (uz); Irish (ga); Belarusian (be); Malagasy (mg); Cebuano (ceb); Haitian (ht); Waray-Waray (war); $^*$: Translation system not available}
        \label{tab:wit_both}
        \end{table*}

        Even with far less data than AT+MBT, \muralbase trained on CC12M+EOBT (5) has much stronger zero-shot performance than M3P (1). With fine-tuning, \muralbase (CC12M+EOBT) improves on both fine-tuned M3P and UC2 (14 vs 10,11), except for Japanese. Though MURAL benefits from four times more image-text pairs than the others (CC12m $>$ CC3M), both M3P and UC2 are more complex cross-encoder models that require other resources. M3P uses several different losses and it relies on a synthetic code-switched data generation process and a pretrained Faster-RCN model to obtain object bounding boxes and labels. MURAL is simpler: it is a dual encoder using just two loss types, and it works directly on raw text and pixels.

        The \textit{translate-train} strategy works well compared to using only multilingual image-text pairs (2 vs 4; 12 vs 13) and versus text-text training (2 vs 6; 12 vs 15). Given this, using translate-train (2) to increase language diversity in image-text pairs \textit{combined} with text-text pair training (6) may yield even more gains. As a zero-shot strategy, \textit{translate-test} also works well . This suggests that SMALR's combination of multilingual encoding and translate-test \cite{burns2020eccv} may improve zero-shot performance further with MURAL (i.e., 3+6+SMALR).
        
        Like others before, we find that training larger models on data of this scale produces remarkable gains: \murallarge obtains big improvements even over \muralbase. \murallarge's results are state-of-the-art for all languages except English (where the larger, English-only \alignhuge is best). \murallarge does this while--as a dual encoder--also supporting efficient retrieval. This makes a huge difference when retrieving from billions of items rather than the 1k to 5k examples of Multi30k's and MS-COCO's test sets (for which expensive, exhaustive comparisons can be performed with cross-encoders). See \citet{geigle-etal-retrieval} for extensive discussion and experiments around the computational cost of cross-encoders versus dual encoders for retrieval.

        \paragraph{Wikipedia Image Text Results.} We extracted two subsets of WIT for evaluation: 1) well-resourced languages and 2) under-resourced languages (more details in Appendix \ref{appendix:wit}). There are no prior results; here, we compare MURAL with \alignmling and \alignen using the translate-test baseline. Table \ref{tab:wit_both} shows \muralbase achieves slightly better zero-shot performance compared to \alignmling on well-resourced languages, and a large boost on the under-represented ones. These results confirm our hypothesis of combining two tasks to address data scarcity in cross modal pairs. For WIT, \murallarge again shows that increasing model capacity improves zero-shot performance dramatically (row 7).
        
        \begin{table}
        \scalebox{.68}{
              \begin{tabular}{ll|ccccccc}
                    & Model & it & es & ru & zh & pl & tr & ko \\
                    \hline
                    -- & mUSE+M3L & 78.9 & 76.7 & 73.6 & 76.1 & 71.7 & 70.9 & 70.7 \\
                    (4) & \alignmling  & 87.9 & 88.8 & 82.3 & 86.5 & 79.8 & 73.5 & 76.6 \\
                    (6) & \muralbase &  88.4 & 89.6 & 83.6 & 88.3 & 86.1 & 84.8 & 82.4 \\
                    (7) & \murallarge & \bf 91.8 & \bf 92.9 & \bf 87.2 & \bf 89.7 & \bf 91.0 & \bf 89.5 & \bf 88.1\\
                    %& ALIGN  & \textbf{88.8} & \textbf{89.5} & \textbf{82.2} & \textbf{59.8} & 78.0 & 75.0 & 12.9 \\
                    %& MURAL &  87.6 & \textbf{89.4} & 80.5 & 57.9 & \textbf{82.6} & \textbf{82.3} & 1.8 \\            
              \end{tabular}
          }
          \centering
        \caption{XTD zero-shot Text{$\rightarrow$}Image Recall@10. }
        \label{tab:xtd10}
        \end{table}
        
        With WIT, the translate-test strategy again proves effective (row 3). It is comparable to both \muralbase and \alignmling in a zero-shot setting-- each wins some contests. Nevertheless, translate-test fails for the extremely under-resourced Waray-Waray language because the NMT system lacks support for it. In all, we found that 27 of WIT's 108 languages lacked NMT support. Thus, we cannot fully rely on translation systems for many under-represented languages; this further bolsters exploration into pivoting on images to overcome data scarcity. Furthermore, simple dual-encoder models are fast and simple at test-time, and thus scale better than translate-test. 
        
        \begin{table*}[t]
            \centering
            \scalebox{0.65}{
            \begin{tabular}{ll|cccc|cccc|cccc|cccc}
                & & \multicolumn{4}{c|}{Image $\rightarrow$ Text}  & \multicolumn{4}{c}{Text $\rightarrow$ Image} & \multicolumn{4}{c}{Text $\rightarrow$ Text} & \multicolumn{4}{c}{Image $\rightarrow$ Image}\\
                & Model & R@1 & R@5 & R@10 & avg r & R@1 & R@5 & R@10 & avg r & R@1 & R@5 & R@10 & avg r & R@1 & R@5 & R@10 & avg r\\
                \hline
                (22) & \cxcbest & 55.9 & 84.2 & 91.8 & - & 41.7 & 72.3 & 83.0 & - & 42.4 & 64.9 & 74.0 & - & 38.5 & 73.6 & 84.9 & - \\
                (13) & \alignmling & 67.1 & 89.0 & 94.2 & 3.6 & 50.0 & 77.3 & 85.9 & 11.5 & 43.5 & 64.7 & 73.5 & 45.4 & 42.6 & 76.6 & 86.2 & 16.0  \\
                (15) & \muralbase & 65.8 & 89.1 & 94.3 & 3.2 & 49.7 & 77.5 & 86.0 & 11.0 & 43.9 & 64.9 & 73.9 & \bf 44.9 & 43.9 & 76.7 & 86.5 & 16.1 \\
                (16) & \murallarge & 74.6 & 92.8 & 96.6 & \bf 2.3 & 57.8 & 83.1 & 90.0 & \bf \pz9.4 & \bf 46.5 & \bf 67.5 & \bf 76.1 & 47.8 & \bf 50.3 & \bf 81.8 & \bf 90.1 & \bf 12.4\\
                (17) & \alignhuge &  \bf 78.1 & \bf 94.3 & \bf 97.4 & - & \bf 61.8 & \bf 84.9 & \bf 91.1 & - & 45.4 & 66.8 & 75.2 & - & 49.4 & 81.4 & 89.1 & - \\
            \end{tabular}
          }
        \caption{CxC Image{$\leftrightarrow$}text (left), Text{$\rightarrow$}Text (middle), and  Image{$\rightarrow$}Image (right) retrieval results. \cxcbest is the strongest model of \citet{parekh-etal-2021-crisscrossed}. \cxcbest and \alignhuge are fine-tuned on MSCOCO data, while \alignmling, \muralbase, and \murallarge are fine-tuned on both Multi30K and MSCOCO data).}
        \label{tab:cxc_retrieval}
        \end{table*}
        
        Finally, both \alignmling and MURAL models benefit from fine-tuning on in-domain multilingual image-text training pairs,\footnote{We fine-tune on WIT training split for 300K steps with initial learning rate 1e-4. Other hyper-parameters are the same as pre-training.} when available; both obtain very large gains across all languages, and also easily beat the translate-test baseline fine-tuned on WIT-en (18, 19, 20 vs 21).

        \textbf{XTD.} As shown in Table \ref{tab:xtd10}, both ALIGN and MURAL obtain massive gains over the best strategy reported by \citet{aggarwal2020towards}---mUSE \cite{yang2019multilingual} with a multimodal metric loss (M3L). \murallarge shows especially strong performance across all languages. Note that we only obtained these scores after all experimentation was done on other datasets---this is methodologically important as there is neither training data nor development data for XTD.

        \begin{table}[t]
        \scalebox{0.78}{
            \centering
            \begin{tabular}{ll|ccc}
                & \multirow{2}{*}{Model} & STS & SIS &  SITS \\
                & & avg $\pm$ std & avg $\pm$ std & avg $\pm$ std\\
                \hline
                (22) & \cxcbest & \bf 74.5 $\pm$ 0.4 & 74.5 $\pm$ 0.9 & 61.9 $\pm$ 1.3 \\
                (13) & \alignmling & 72.7 $\pm$ 0.4 & \bf 80.4 $\pm$ 0.7 & 63.7 $\pm$ 1.3 \\
                (15) & \muralbase & 73.9 $\pm$ 0.4 & 80.0 $\pm$ 0.7 & 64.0 $\pm$ 1.2  \\
                (16) & \murallarge & 74.1 $\pm$ 0.4 & \bf 80.4 $\pm$ 0.7 &  67.1 $\pm$ 1.3  \\
                (17) & \alignhuge  & 72.9  $\pm$ 0.4 & 77.2 $\pm$ 0.8 & \bf 67.6 $\pm$ 1.2 \\
            \end{tabular}
          }
        \caption{Semantic Simliarity using CxC.}
        \label{tab:cxc_correlation}
        \end{table}

        \textbf{Crisscrossed Captions.} 
        For CxC image-text retrieval (Table \ref{tab:cxc_retrieval}), \alignhuge scores highest across all metrics; it is the largest model and was trained only on English Alt-Text. \alignmling also beats \muralbase for image-text retrieval, but the latter comes back with better text-text and image-image scores. This indicates that MURAL's text-text task balances both encoders better than a loss focused only on image-text pairs. Similarly, \murallarge beats \alignhuge for both text-text and image-image retrieval, despite the fact that \alignhuge uses a much larger image encoder.
        
        %Both \muralbase and \muralbase achieve a drastic drop in Average Rank compared to \alignmling indicating that on average the retrieved co-captions are ranked lower.
        %In image-image similarity comparison, both \muralbase and \alignhuge get strong results compared to other models. This indicates that using larger image encoders are important for image retrieval. 
        
        The correlation results given in Table \ref{tab:cxc_correlation} tell an interesting story. Contrary to intuition and retrieval results,  Semantic Image Similarity (SIS) seems connected with multilinguality, as all Alt-Text models (\alignmling, \muralbase, \murallarge) perform nearly the same (and better).  \cxcbest scores the highest on Semantic Text Similarity (STS) followed closely by \murallarge. It is worth noting that \cxcbest was trained with MSCOCO co-captions which could explain its high correlation.
        % It is interesting to note the models are leaning towards either the retrieval metrics or correlation metrics for text-text similarity. \murallarge has the best text-text retrieval, but lowest on STS correlation (Table \ref{tab:cxc_retrieval}, row 16). 
        Semantic Image-Text Similarity (SITS) agrees with Image-Text retrieval results the most, with both \murallarge and \alignhuge performing considerably better than others. However, with the SITS metric, the gap between both these models diminishes, indicating that \alignhuge is probably more focused on getting positive matches while \murallarge captures non-matches more effectively.
        
        %Comparing \alignmling with \muralbase, we see similar performance which hints that adding text-text loss does not impact image-image similarity much.
        
        The combined retrieval and correlation lens of CxC indicates there is much more to evaluating multimodal representations than the predominant cross-modal retrieval tasks. Ranking a set of items in a manner consistent with human similarity judgments is arguably a harder task than getting a single paired item to be more similar than nearly all others. These two perspectives may reveal useful tensions in finer-grained semantic distinctions. In fact, it is with these correlation measures that we expect cross-encoders to shine compared to the retrieval-oriented dual encoders.

\section{Analysis}

% \begin{figure*}[t]
%     \centering
    
%     \subfloat{\includegraphics[width=0.47\linewidth]{emnlp2020-templates/images/embed_labse.png} }
%     \qquad
%     \subfloat{{\includegraphics[width=0.47\linewidth]{emnlp2020-templates/images/embed_mural.png} }}
%     \caption{Left visualization plots LaBSE embeddings. Right visualization plots MURAL embeddings.}
%     \label{fig:embed_visualization}
% \end{figure*}

\begin{figure*}[t]
    \centering
    \begin{subfigure}[b]{0.47\textwidth}
         \centering
         \includegraphics[width=\textwidth]{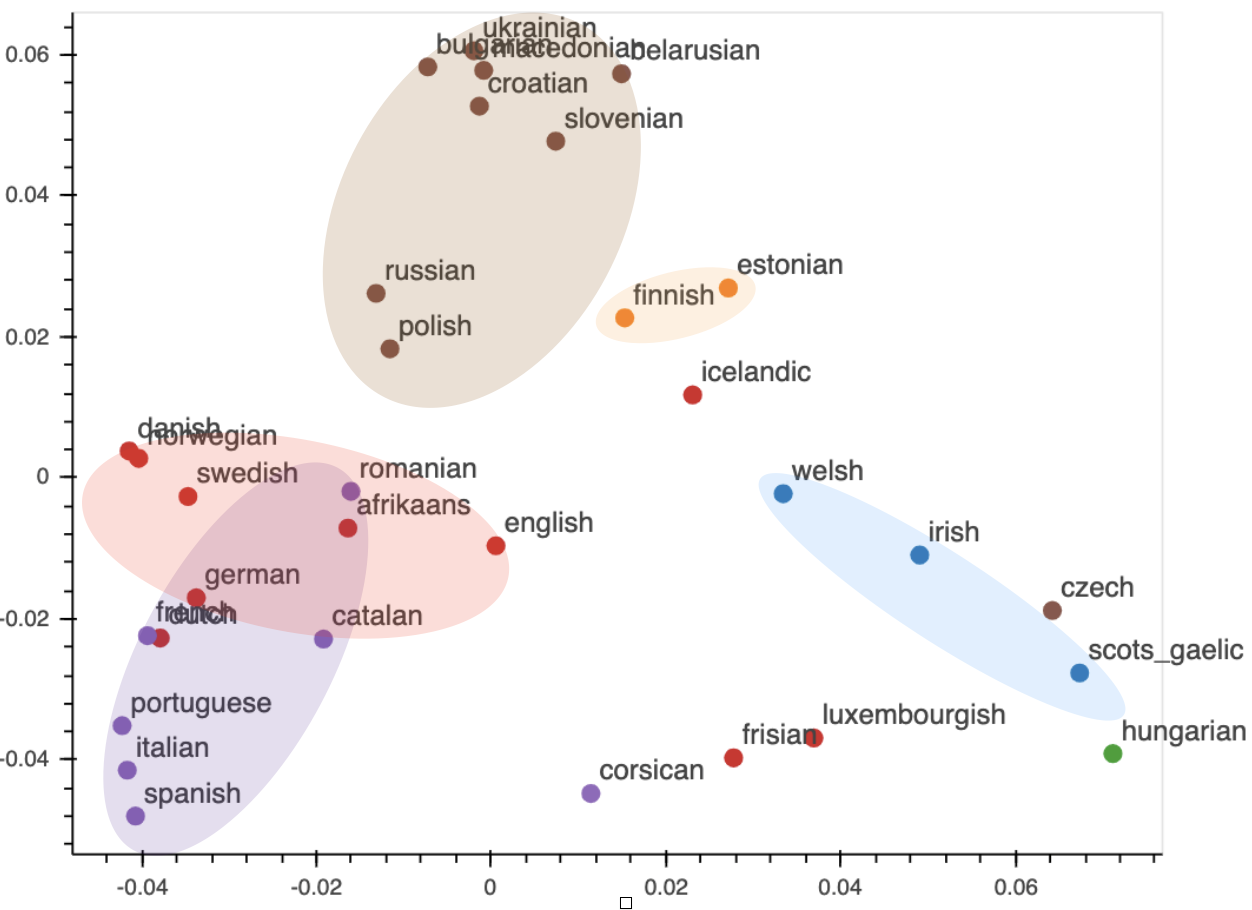}
         \caption{LaBSE representations}
         \label{fig:y equals x}
     \end{subfigure}
    % \hfill
    \begin{subfigure}[b]{0.47\textwidth}
         \centering
         \includegraphics[width=\textwidth]{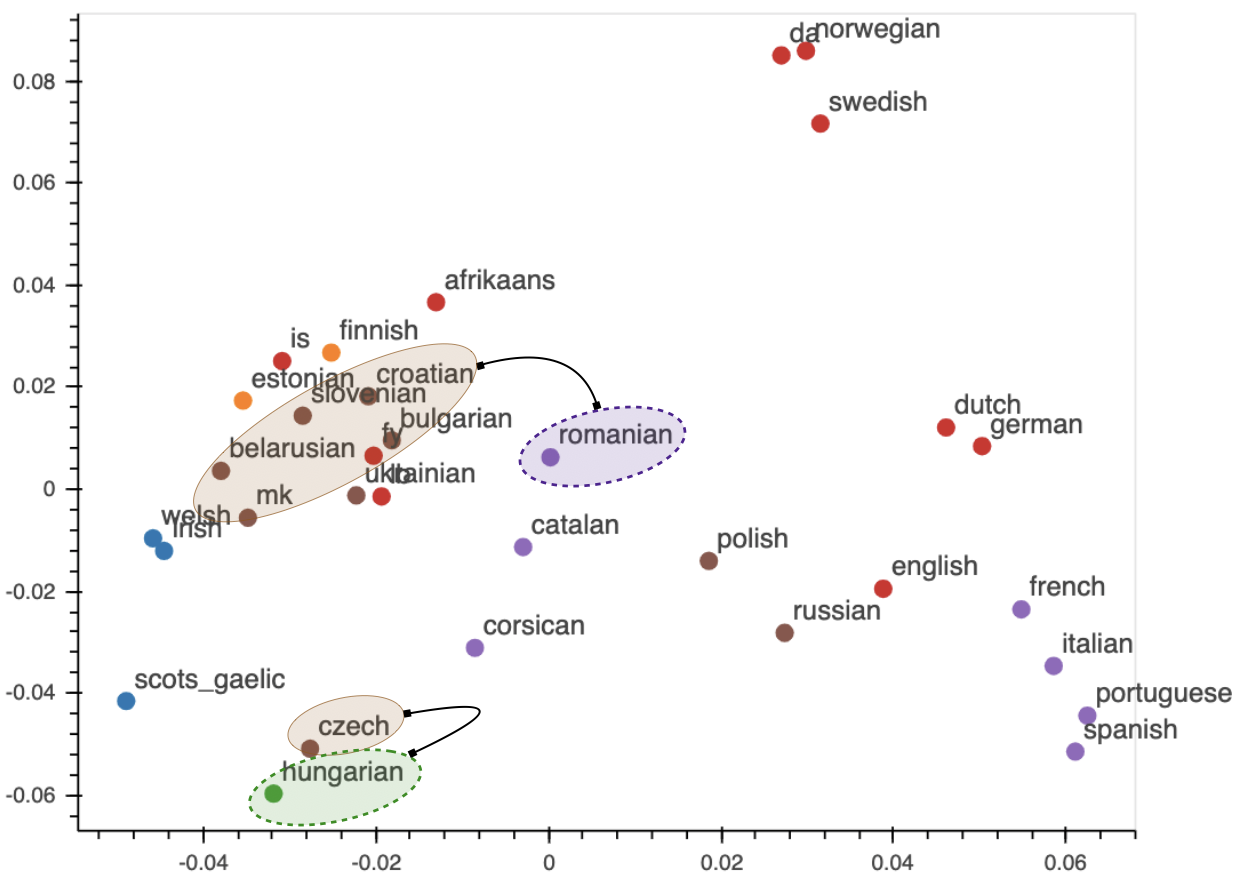}
         \caption{\muralbase representations}
         \label{fig:y equals x}
    \end{subfigure}
     
    \caption{Visualization of text representations of LaBSE \cite{feng2020language} and MURAL for 35 languages using laplacian eigen values and SVCCA scores. Languages are color coded based on their genealogical association.}
    \label{fig:embed_visualization}
\end{figure*}

\textbf{Embedding Visualization}. 
We visualize multilingual text representations using Singular Value
Canonical Correlation Analysis (SVCCA) \citep{raghu2017svcca}, which allows similarity scores to be computed between languages. Using SVCCA scores computed for 100 languages, we plot a 2-dimensional visualization using Laplacian Eigenmaps \cite{Belkin2003LaplacianEF}. Following \citet{kudugunta2019investigating}, we do so for a subset of languages belonging to the Germanic, Romance, Slavic, Uralic, Finnic, Celtic, and Finno-Ugric language families (widely spoken in Europe and Western Asia). For a fair evaluation, we artificially create a multilingual aligned dataset by using Google's Translation system to translate 1K English captions from the Multi30K dataset to 100 languages.

Figure \ref{fig:embed_visualization} plots the embedding in a 2-dimensional space for two models: 1) LaBSE, a multilingual \textit{text-only} sentence representation model \cite{feng2020language} and 2) MURAL, a multingual \textit{multimodal} model. It is evident from the visualization of LaBSE representations that embeddings group largely based on genealogical connections between languages, a phenomenon observed previously in \citet{kudugunta2019investigating}. In addition to groupings informed by linguistic genealogy, the MURAL visualization interestingly shows some clusters which are in line with areal linguistics and contact linguistics. Notably, Romanian (ro) is closer to the Slavic languages like Bulgarian (bg), Macedonian (mk) in MURAL than it is for LaBSE, which is in line with the Balkan Sprachbund \cite{joseph1999romanian}. English (en) and French (fr) are also embedded closer to each other, reflecting their extensive contact \cite{haeberli2014english}. Another possible language contact brings Finnic languages, Estonian (et) and Finnish (fi), closer to the Slavic languages cluster.

The fact that MURAL pivots on images as well as translations thus appears to add an additional view on language relatedness as learned in deep representations, beyond the language family clustering observed in a text-only setting. This suggests potential future work to explore different linguistic phenomena in these representations. It also suggests that it may be worth trying to improve multimodal, multilingual representations for a given lower-resource language by pivoting on a well-resourced language that is linguistically related or which has been in significant contact with it--similar to previous studies for machine translation \cite{islam2013source}.

% What had been written
%This opens way for potential future work to explore different linguistics phenomena and also to use these inferences in improving multimodal multilingual representations pivoting on a more linguistically informed closer well-resourced language--similar to what is being explored for machine translation \cite{islam2013source}.

\begin{figure}[t]
    \centering
    \centering
    \includegraphics[width=0.47\textwidth]{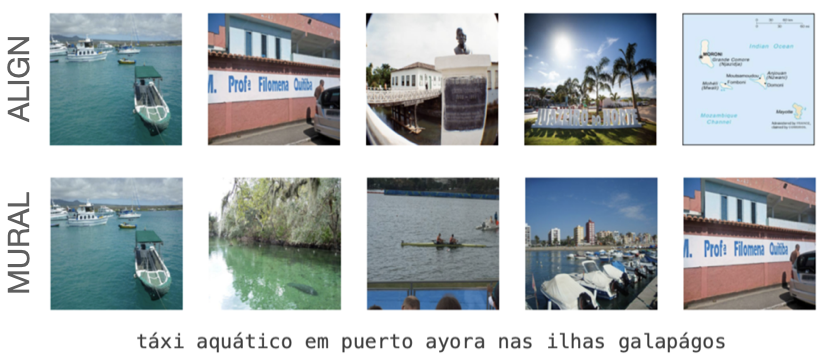}
    \caption{Portuguese: retrieval coherence. (\textit{``Water taxi in Puerto Ayora in the Galapágos Islands.''})}
    \label{fig:pt_fidelity}
\end{figure}

\begin{figure}[t]
    \centering
    \centering
    \includegraphics[width=0.48\textwidth]{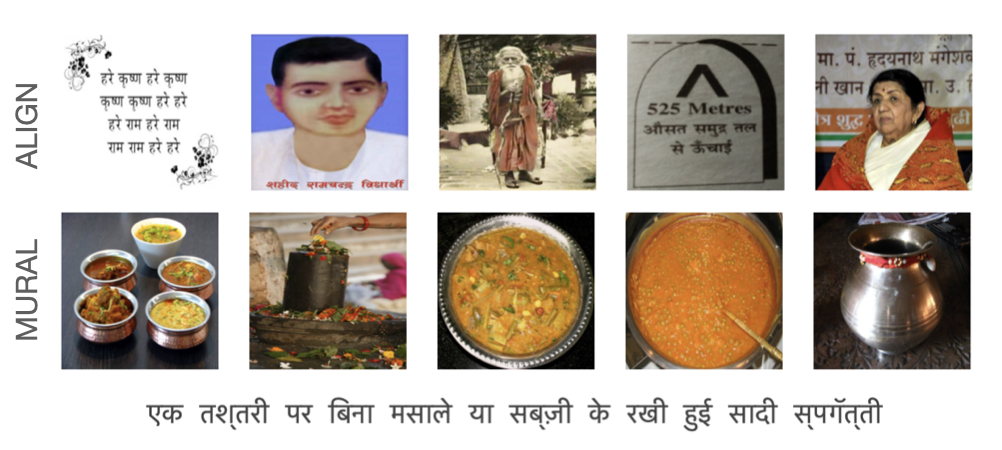}
     
    \caption{Hindi: Text{$\rightarrow$}Image. (\textit{``A bowl containing plain noodles without any spices or vegetables.''})}
    \label{fig:hindi_culture}
\end{figure}

\begin{figure}[t]
    \centering
    \centering
    \begin{subfigure}[b]{0.1\textwidth}
         \includegraphics[width=\textwidth]{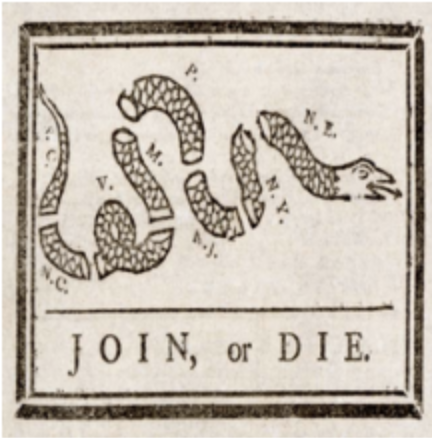}
     \end{subfigure}
    \begin{subfigure}[b]{0.1\textwidth}
         \includegraphics[width=\textwidth]{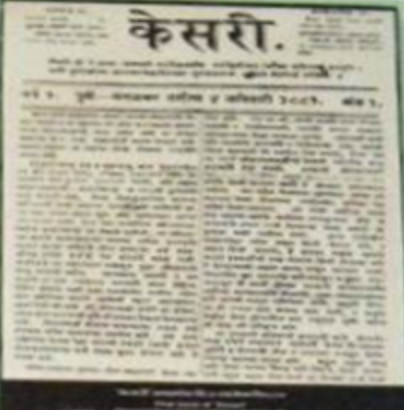}
     \end{subfigure}
     \begin{subfigure}[b]{0.1\textwidth}
         \includegraphics[width=\textwidth]{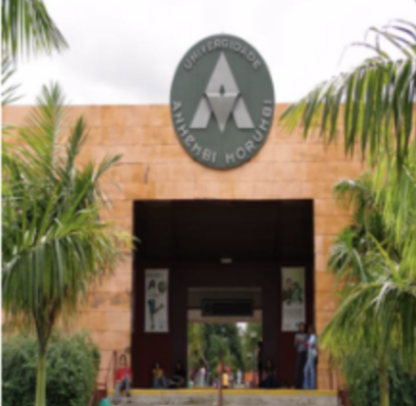}
     \end{subfigure}
     \begin{subfigure}[b]{0.1\textwidth}
         \includegraphics[width=\textwidth]{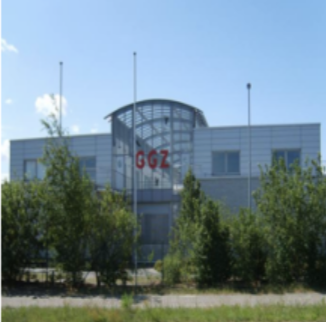}
     \end{subfigure}
    
    \caption{Image $\rightarrow$ Text examples where recognizing text in the input image would greatly help.}
    \label{fig:ocr}
\end{figure}

\textbf{Retrieval Error Analysis}. We analyzed zero-shot retrieved examples on WIT for \alignmling and \muralbase for English (en), Hindi (hi), French (fr), and Portugese (pt). We list some examples here that indicate the value of using translation pair data for learning multilingual multimodal representations. See Appendix \ref{appendix:error_analysis} for more examples.

Across languages, for both Image$\rightarrow$Text retrieval and Text$\rightarrow$Image, we observed that MURAL displays better fidelity to the concepts described in the image and text. For instance, in Fig. \ref{fig:pt_fidelity} ALIGN's top five results are somewhat scattered, whereas MURAL's results cohere better around boats with people (water taxis) near land (islands).

For under-resourced languages like Hindi, MURAL shows an improvement with respect to retrieving results that are culturally more suited to the language (Fig. \ref{fig:hindi_culture}).

Finally, with both models, retrieval for some examples could greatly benefit from better recognition of words present in the images. Fig. \ref{fig:ocr} shows examples where extracting text from the images would make Image{$\rightarrow$}Text almost trivial.

\section{Conclusion}

English provides a strong starting point for learning multilingual representations because it is so wide-spread and examples of English paired with other languages can be gathered well-beyond that of any other language, currently. We exploit this to train on translation pairs as a means to improve handling of multilingual inputs in cross-modal representations. With simple dual encoder models trained on large-scale datasets via contrastive learning, we obtain consistent, strong retrieval performance across all languages---especially under-resourced ones. Our error analysis also indicates that this helps increasing cultural specificity and diversity of the retrieved examples. The nuanced results we obtained for CxC also indicate that further improvements in such models might come from better calibration of the different tasks during learning. We also expect that more aggressive use of the \textit{translate-train} strategy will straightforwardly yield further gains.

Embedding visualizations of MURAL's text representations also illustrates how languages cluster based on multimodal learning. Prior work has shown that English is not the ideal pivot language for many under-resourced languages \cite{mulcaire2019polyglot, lample2019cross}. Our improvements for multilingual and multimodal models suggest further investigations into which well-resourced languages can be better pivots for learning representations for under-resourced languages. In addition to reflecting established language groupings, it also opens up possibilities of discovering new clusters. For instance, the proximity of Hungarian and Czech (Fig \ref{fig:embed_visualization}) for MURAL might be attributed to the geographical proximity of these languages, and warrants further analysis.

\section{Ethics}

Models trained on data collected from the web show strong results, and we are particularly encouraged by the fact that doing so leads to large improvements on under-resourced languages---and does so without requiring large amounts of (or any) image-text training data for those languages. Nevertheless, we should take utmost caution when using large datasets which went through minimal filtering processes. There could be potential biases both in the training data and models trained on them. Conscious research efforts should be made to detect and address such biases prior to releasing and using these models.

Fortunately, with prior research work in ethical AI research, it is possible to use findings from these areas to make the cross-modal models more accountable for their retrieval and broader use. We believe our findings and models can contribute positively to better understanding issues of and opportunities for addressing ethics, fairness, bias, and responsibility--especially with respect to cross-cultural issues--in language and images.\
\section*{Acknowledgments}
We thank the anonymous reviewers for their helpful feedback. We also thank Anosh Raj and Daphne Luong from Google for initial reviews of the manuscript. We thank Zarana Parekh for helping us setup evaluation on the CxC dataset, Orhan Firat for providing guidance on vocabulary coverage, Yuqing Chen and Apu Shah for assisting with latency metrics of the NMT models.\
\bibliography{emnlp2021}

\begin{thebibliography}{54}
\expandafter\ifx\csname natexlab\endcsname\relax\def\natexlab#1{#1}\fi

\bibitem[{Aggarwal and Kale(2020)}]{aggarwal2020towards}
Pranav Aggarwal and Ajinkya Kale. 2020.
\newblock \href {https://arxiv.org/abs/2012.05107} {Towards zero-shot
  {C}ross-lingual {I}mage retrieval}.
\newblock \emph{arXiv preprint arXiv:2012.05107}.

\bibitem[{Agi{\'c} and Vuli{\'c}(2019)}]{agic2020jw300}
{\v{Z}}eljko Agi{\'c} and Ivan Vuli{\'c}. 2019.
\newblock \href {https://doi.org/10.18653/v1/P19-1310} {{JW}300: A
  wide-coverage parallel corpus for low-resource languages}.
\newblock In \emph{Proceedings of the 57th Annual Meeting of the Association
  for Computational Linguistics}, pages 3204--3210, Florence, Italy.
  Association for Computational Linguistics.

\bibitem[{Barrault et~al.(2018)Barrault, Bougares, Specia, Lala, Elliott, and
  Frank}]{barrault2018findings}
Lo{\"\i}c Barrault, Fethi Bougares, Lucia Specia, Chiraag Lala, Desmond
  Elliott, and Stella Frank. 2018.
\newblock \href {https://doi.org/10.18653/v1/W18-6402} {Findings of the third
  shared task on multimodal machine translation}.
\newblock In \emph{Proceedings of the Third Conference on Machine Translation:
  Shared Task Papers}, pages 304--323, Belgium, Brussels. Association for
  Computational Linguistics.

\bibitem[{Belkin and Niyogi(2003)}]{Belkin2003LaplacianEF}
Mikhail Belkin and P.~Niyogi. 2003.
\newblock Laplacian eigenmaps for dimensionality reduction and data
  representation.
\newblock \emph{Neural Computation}, 15:1373--1396.

\bibitem[{Burns et~al.(2020)Burns, Kim, Wijaya, Saenko, and
  Plummer}]{burns2020eccv}
Andrea Burns, Donghyun Kim, Derry Wijaya, Kate Saenko, and Bryan~A. Plummer.
  2020.
\newblock Learning to scale multilingual representations for vision-language
  tasks.
\newblock In \emph{The European Conference on Computer Vision (ECCV)}.

\bibitem[{Changpinyo et~al.(2021)Changpinyo, Sharma, Ding, and
  Soricut}]{changpinyo2021conceptual}
Soravit Changpinyo, Piyush Sharma, Nan Ding, and Radu Soricut. 2021.
\newblock \href {https://arxiv.org/abs/2102.08981} {Conceptual 12{M}: Pushing
  web-scale image-text pre-training to recognize long-tail visual concepts}.
\newblock \emph{arXiv preprint arXiv:2102.08981}.

\bibitem[{Chen et~al.(2020)Chen, Kornblith, Norouzi, and
  Hinton}]{chen2020simple}
Ting Chen, Simon Kornblith, Mohammad Norouzi, and Geoffrey~E. Hinton. 2020.
\newblock \href {http://proceedings.mlr.press/v119/chen20j.html} {A simple
  framework for contrastive learning of visual representations}.
\newblock In \emph{Proceedings of the 37th International Conference on Machine
  Learning, {ICML} 2020, 13-18 July 2020, Virtual Event}, volume 119 of
  \emph{Proceedings of Machine Learning Research}, pages 1597--1607. {PMLR}.

\bibitem[{Conneau and Lample(2019)}]{lample2019cross}
Alexis Conneau and Guillaume Lample. 2019.
\newblock \href
  {https://proceedings.neurips.cc/paper/2019/hash/c04c19c2c2474dbf5f7ac4372c5b9af1-Abstract.html}
  {Cross-lingual language model pretraining}.
\newblock In \emph{Advances in Neural Information Processing Systems 32: Annual
  Conference on Neural Information Processing Systems 2019, NeurIPS 2019,
  December 8-14, 2019, Vancouver, BC, Canada}, pages 7057--7067.

\bibitem[{Devlin et~al.(2019)Devlin, Chang, Lee, and
  Toutanova}]{devlin-etal-2019-bert}
Jacob Devlin, Ming-Wei Chang, Kenton Lee, and Kristina Toutanova. 2019.
\newblock \href {https://doi.org/10.18653/v1/N19-1423} {{BERT}: Pre-training of
  deep bidirectional transformers for language understanding}.
\newblock In \emph{Proceedings of the 2019 Conference of the North {A}merican
  Chapter of the Association for Computational Linguistics: Human Language
  Technologies, Volume 1 (Long and Short Papers)}, pages 4171--4186,
  Minneapolis, Minnesota. Association for Computational Linguistics.

\bibitem[{Elliott et~al.(2017)Elliott, Frank, Barrault, Bougares, and
  Specia}]{elliott2017findings}
Desmond Elliott, Stella Frank, Lo{\"\i}c Barrault, Fethi Bougares, and Lucia
  Specia. 2017.
\newblock \href {https://doi.org/10.18653/v1/W17-4718} {Findings of the second
  shared task on multimodal machine translation and multilingual image
  description}.
\newblock In \emph{Proceedings of the Second Conference on Machine
  Translation}, pages 215--233, Copenhagen, Denmark. Association for
  Computational Linguistics.

\bibitem[{Elliott et~al.(2016)Elliott, Frank, Sima{'}an, and
  Specia}]{elliott2016multi30k}
Desmond Elliott, Stella Frank, Khalil Sima{'}an, and Lucia Specia. 2016.
\newblock \href {https://doi.org/10.18653/v1/W16-3210} {{M}ulti30{K}:
  Multilingual {E}nglish-{G}erman image descriptions}.
\newblock In \emph{Proceedings of the 5th Workshop on Vision and Language},
  pages 70--74, Berlin, Germany. Association for Computational Linguistics.

\bibitem[{Espl{\`a} et~al.(2019)Espl{\`a}, Forcada, Ram{\'\i}rez-S{\'a}nchez,
  and Hoang}]{espla-etal-2019-paracrawl}
Miquel Espl{\`a}, Mikel Forcada, Gema Ram{\'\i}rez-S{\'a}nchez, and Hieu Hoang.
  2019.
\newblock \href {https://aclanthology.org/W19-6721} {{P}ara{C}rawl: Web-scale
  parallel corpora for the languages of the {EU}}.
\newblock In \emph{Proceedings of Machine Translation Summit XVII: Translator,
  Project and User Tracks}, pages 118--119, Dublin, Ireland. European
  Association for Machine Translation.

\bibitem[{Feng et~al.(2020)Feng, Yang, Cer, Arivazhagan, and
  Wang}]{feng2020language}
Fangxiaoyu Feng, Yinfei Yang, Daniel Cer, Naveen Arivazhagan, and Wei Wang.
  2020.
\newblock \href {https://arxiv.org/abs/2007.01852} {Language-agnostic {BERT}
  sentence embedding}.
\newblock \emph{arXiv preprint arXiv:2007.01852}.

\bibitem[{Geigle et~al.(2021)Geigle, Pfeiffer, Reimers, Vulic, and
  Gurevych}]{geigle-etal-retrieval}
Gregor Geigle, Jonas Pfeiffer, Nils Reimers, Ivan Vulic, and Iryna Gurevych.
  2021.
\newblock \href {http://arxiv.org/abs/2103.11920} {Retrieve fast, rerank smart:
  Cooperative and joint approaches for improved cross-modal retrieval}.
\newblock \emph{CoRR}, abs/2103.11920.

\bibitem[{Gella et~al.(2017)Gella, Sennrich, Keller, and
  Lapata}]{gella2017image}
Spandana Gella, Rico Sennrich, Frank Keller, and Mirella Lapata. 2017.
\newblock \href {https://doi.org/10.18653/v1/D17-1303} {Image pivoting for
  learning multilingual multimodal representations}.
\newblock In \emph{Proceedings of the 2017 Conference on Empirical Methods in
  Natural Language Processing}, pages 2839--2845, Copenhagen, Denmark.
  Association for Computational Linguistics.

\bibitem[{Groenewald and du~Plooy(2010)}]{groenewald2010processing}
Hendrik~J Groenewald and Liza du~Plooy. 2010.
\newblock Processing parallel text corpora for three {S}outh {A}frican language
  pairs in the {A}utshumato project.
\newblock \emph{AfLaT 2010}, page~27.

\bibitem[{Guo et~al.(2019)Guo, Pasunuru, and Bansal}]{guo2019autosem}
Han Guo, Ramakanth Pasunuru, and Mohit Bansal. 2019.
\newblock \href {https://doi.org/10.18653/v1/N19-1355} {{A}uto{S}e{M}:
  Automatic task selection and mixing in multi-task learning}.
\newblock In \emph{Proceedings of the 2019 Conference of the North {A}merican
  Chapter of the Association for Computational Linguistics: Human Language
  Technologies, Volume 1 (Long and Short Papers)}, pages 3520--3531,
  Minneapolis, Minnesota. Association for Computational Linguistics.

\bibitem[{Guo et~al.(2020)Guo, Sun, Lindgren, Geng, Simcha, Chern, and
  Kumar}]{avq_2020}
Ruiqi Guo, Philip Sun, Erik Lindgren, Quan Geng, David Simcha, Felix Chern, and
  Sanjiv Kumar. 2020.
\newblock \href {http://proceedings.mlr.press/v119/guo20h.html} {Accelerating
  large-scale inference with anisotropic vector quantization}.
\newblock In \emph{Proceedings of the 37th International Conference on Machine
  Learning, {ICML} 2020, 13-18 July 2020, Virtual Event}, volume 119 of
  \emph{Proceedings of Machine Learning Research}, pages 3887--3896. {PMLR}.

\bibitem[{Guzm\'{a}n et~al.(2019)Guzm\'{a}n, Chen, Ott, Pino, Lample, Koehn,
  Chaudhary, and Ranzato}]{flores_dataset}
Francisco Guzm\'{a}n, Peng-Jen Chen, Myle Ott, Juan Pino, Guillaume Lample,
  Philipp Koehn, Vishrav Chaudhary, and Marc'Aurelio Ranzato. 2019.
\newblock \href {https://arxiv.org/abs/1902.01382} {Two new evaluation datasets
  for low-resource machine translation: {N}epali-{E}nglish and
  {S}inhala-{E}nglish}.

\bibitem[{Haddow and Kirefu(2020)}]{haddow2020pmindia}
Barry Haddow and Faheem Kirefu. 2020.
\newblock \href {https://arxiv.org/abs/2001.09907} {{PMIndia}--a collection of
  parallel corpora of languages of {I}ndia}.
\newblock \emph{arXiv preprint arXiv:2001.09907}.

\bibitem[{Haeberli(2014)}]{haeberli2014english}
Eric Haeberli. 2014.
\newblock When {E}nglish meets {F}rench: A case study of language contact in
  {M}iddle {E}nglish.
\newblock \emph{Papers Dedicated to Jacques Moeschler}.

\bibitem[{Hernandez and Nguyen(2020)}]{hernandez2020ubiqus}
Fran{\c{c}}ois Hernandez and Vincent Nguyen. 2020.
\newblock \href {https://aclanthology.org/2020.wmt-1.21} {The ubiqus
  {E}nglish-{I}nuktitut system for {WMT}20}.
\newblock In \emph{Proceedings of the Fifth Conference on Machine Translation},
  pages 213--217, Online. Association for Computational Linguistics.

\bibitem[{Islam and Hoenen(2013)}]{islam2013source}
Zahurul Islam and Armin Hoenen. 2013.
\newblock \href {https://aclanthology.org/I13-1185} {Source and translation
  classification using most frequent words}.
\newblock In \emph{Proceedings of the Sixth International Joint Conference on
  Natural Language Processing}, pages 1299--1305, Nagoya, Japan. Asian
  Federation of Natural Language Processing.

\bibitem[{Jia et~al.(2021)Jia, Yang, Xia, Chen, Parekh, Pham, Le, Sung, Li, and
  Duerig}]{jia2021scaling}
Chao Jia, Yinfei Yang, Ye~Xia, Yi-Ting Chen, Zarana Parekh, Hieu Pham, Quoc~V
  Le, Yunhsuan Sung, Zhen Li, and Tom Duerig. 2021.
\newblock \href {https://arxiv.org/abs/2102.05918} {Scaling up visual and
  vision-language representation learning with noisy text supervision}.
\newblock \emph{arXiv preprint arXiv:2102.05918}.

\bibitem[{Joseph(1999)}]{joseph1999romanian}
Brian~D Joseph. 1999.
\newblock Romanian and the {B}alkans: Some comparative perspectives.
\newblock \emph{The Emergence of the Modern Language Sciences. Studies on the
  Transition from Historical-Comparative to Structural Linguistics in Honour of
  EFK Koerner}, 2:218--235.

\bibitem[{Karpathy and Li(2015)}]{karpathy2015deep}
Andrej Karpathy and Fei{-}Fei Li. 2015.
\newblock \href {https://doi.org/10.1109/CVPR.2015.7298932} {Deep
  visual-semantic alignments for generating image descriptions}.
\newblock In \emph{{IEEE} Conference on Computer Vision and Pattern
  Recognition, {CVPR} 2015, Boston, MA, USA, June 7-12, 2015}, pages
  3128--3137. {IEEE} Computer Society.

\bibitem[{Koehn(2005)}]{koehn2005europarl}
Philipp Koehn. 2005.
\newblock \href {https://aclanthology.org/2005.mtsummit-papers.11} {{E}uroparl:
  A parallel corpus for statistical machine translation}.
\newblock In \emph{Proceedings of Machine Translation Summit X: Papers}, pages
  79--86, Phuket, Thailand.

\bibitem[{Kudugunta et~al.(2019)Kudugunta, Bapna, Caswell, and
  Firat}]{kudugunta2019investigating}
Sneha Kudugunta, Ankur Bapna, Isaac Caswell, and Orhan Firat. 2019.
\newblock \href {https://doi.org/10.18653/v1/D19-1167} {Investigating
  multilingual {NMT} representations at scale}.
\newblock In \emph{Proceedings of the 2019 Conference on Empirical Methods in
  Natural Language Processing and the 9th International Joint Conference on
  Natural Language Processing (EMNLP-IJCNLP)}, pages 1565--1575, Hong Kong,
  China. Association for Computational Linguistics.

\bibitem[{Li et~al.(2019)Li, Xu, Wang, Lan, Jia, Yang, and Xu}]{Li2019COCOCNFC}
Xirong Li, Chaoxi Xu, X.~Wang, Weiyu Lan, Zhengxiong Jia, Gang Yang, and
  Jieping Xu. 2019.
\newblock {COCO-CN} for cross-lingual image tagging, captioning, and retrieval.
\newblock \emph{IEEE Transactions on Multimedia}, 21:2347--2360.

\bibitem[{Lin et~al.(2014)Lin, Maire, Belongie, Hays, Perona, Ramanan,
  Doll{\'a}r, and Zitnick}]{lin2014microsoft}
Tsung-Yi Lin, Michael Maire, Serge Belongie, James Hays, Pietro Perona, Deva
  Ramanan, Piotr Doll{\'a}r, and C~Lawrence Zitnick. 2014.
\newblock Microsoft coco: Common objects in context.
\newblock In \emph{European conference on computer vision}, pages 740--755.
  Springer.

\bibitem[{Mulcaire et~al.(2019)Mulcaire, Kasai, and
  Smith}]{mulcaire2019polyglot}
Phoebe Mulcaire, Jungo Kasai, and Noah~A. Smith. 2019.
\newblock \href {https://doi.org/10.18653/v1/N19-1392} {Polyglot contextual
  representations improve crosslingual transfer}.
\newblock In \emph{Proceedings of the 2019 Conference of the North {A}merican
  Chapter of the Association for Computational Linguistics: Human Language
  Technologies, Volume 1 (Long and Short Papers)}, pages 3912--3918,
  Minneapolis, Minnesota. Association for Computational Linguistics.

\bibitem[{Nakazawa et~al.(2016)Nakazawa, Yaguchi, Uchimoto, Utiyama, Sumita,
  Kurohashi, and Isahara}]{Nakazawa2016ASPECAS}
Toshiaki Nakazawa, Manabu Yaguchi, Kiyotaka Uchimoto, Masao Utiyama, Eiichiro
  Sumita, Sadao Kurohashi, and Hitoshi Isahara. 2016.
\newblock \href {https://aclanthology.org/L16-1350} {{ASPEC}: {A}sian
  scientific paper excerpt corpus}.
\newblock In \emph{Proceedings of the Tenth International Conference on
  Language Resources and Evaluation ({LREC}'16)}, pages 2204--2208,
  Portoro{\v{z}}, Slovenia. European Language Resources Association (ELRA).

\bibitem[{Neubig(2011)}]{neubig11kftt}
Graham Neubig. 2011.
\newblock The {Kyoto} free translation task.
\newblock http://www.phontron.com/kftt.

\bibitem[{Ni et~al.(2021)Ni, Huang, Su, Cui, Bharti, Wang, Zhang, and
  Duan}]{huang2020m3p}
Minheng Ni, Haoyang Huang, Lin Su, Edward Cui, Taroon Bharti, Lijuan Wang,
  Dongdong Zhang, and Nan Duan. 2021.
\newblock M3p: Learning universal representations via multitask multilingual
  multimodal pre-training.
\newblock In \emph{Proceedings of the IEEE/CVF Conference on Computer Vision
  and Pattern Recognition}, pages 3977--3986.

\bibitem[{Parekh et~al.(2021)Parekh, Baldridge, Cer, Waters, and
  Yang}]{parekh-etal-2021-crisscrossed}
Zarana Parekh, Jason Baldridge, Daniel Cer, Austin Waters, and Yinfei Yang.
  2021.
\newblock \href {https://aclanthology.org/2021.eacl-main.249} {Crisscrossed
  captions: Extended intramodal and intermodal semantic similarity judgments
  for {MS}-{COCO}}.
\newblock In \emph{Proceedings of the 16th Conference of the European Chapter
  of the Association for Computational Linguistics: Main Volume}, pages
  2855--2870, Online. Association for Computational Linguistics.

\bibitem[{Pryzant et~al.(2018)Pryzant, Chung, Jurafsky, and
  Britz}]{pryzant_jesc_2018}
Reid Pryzant, Youngjoo Chung, Dan Jurafsky, and Denny Britz. 2018.
\newblock \href {https://aclanthology.org/L18-1182} {{JESC}:
  {J}apanese-{E}nglish subtitle corpus}.
\newblock In \emph{Proceedings of the Eleventh International Conference on
  Language Resources and Evaluation ({LREC} 2018)}, Miyazaki, Japan. European
  Language Resources Association (ELRA).

\bibitem[{Radford et~al.(2021)Radford, Kim, Hallacy, Ramesh, Goh, Agarwal,
  Sastry, Askell, Mishkin, Clark et~al.}]{radford2021learning}
Alec Radford, Jong~Wook Kim, Chris Hallacy, Aditya Ramesh, Gabriel Goh,
  Sandhini Agarwal, Girish Sastry, Amanda Askell, Pamela Mishkin, Jack Clark,
  et~al. 2021.
\newblock \href {https://arxiv.org/abs/2103.00020} {Learning transferable
  visual models from natural language supervision}.
\newblock \emph{arXiv preprint arXiv:2103.00020}.

\bibitem[{Raghu et~al.(2017)Raghu, Gilmer, Yosinski, and
  Sohl{-}Dickstein}]{raghu2017svcca}
Maithra Raghu, Justin Gilmer, Jason Yosinski, and Jascha Sohl{-}Dickstein.
  2017.
\newblock \href
  {https://proceedings.neurips.cc/paper/2017/hash/dc6a7e655d7e5840e66733e9ee67cc69-Abstract.html}
  {{SVCCA:} singular vector canonical correlation analysis for deep learning
  dynamics and interpretability}.
\newblock In \emph{Advances in Neural Information Processing Systems 30: Annual
  Conference on Neural Information Processing Systems 2017, December 4-9, 2017,
  Long Beach, CA, {USA}}, pages 6076--6085.

\bibitem[{Schwenk et~al.(2021)Schwenk, Chaudhary, Sun, Gong, and
  Guzm{\'a}n}]{schwenk2019wikimatrix}
Holger Schwenk, Vishrav Chaudhary, Shuo Sun, Hongyu Gong, and Francisco
  Guzm{\'a}n. 2021.
\newblock \href {https://aclanthology.org/2021.eacl-main.115} {{W}iki{M}atrix:
  Mining 135{M} parallel sentences in 1620 language pairs from {W}ikipedia}.
\newblock In \emph{Proceedings of the 16th Conference of the European Chapter
  of the Association for Computational Linguistics: Main Volume}, pages
  1351--1361, Online. Association for Computational Linguistics.

\bibitem[{Sharma et~al.(2018)Sharma, Ding, Goodman, and
  Soricut}]{sharma-etal-2018-conceptual}
Piyush Sharma, Nan Ding, Sebastian Goodman, and Radu Soricut. 2018.
\newblock \href {https://doi.org/10.18653/v1/P18-1238} {Conceptual captions: A
  cleaned, hypernymed, image alt-text dataset for automatic image captioning}.
\newblock In \emph{Proceedings of the 56th Annual Meeting of the Association
  for Computational Linguistics (Volume 1: Long Papers)}, pages 2556--2565,
  Melbourne, Australia. Association for Computational Linguistics.

\bibitem[{Siddhant et~al.(2020)Siddhant, Bapna, Cao, Firat, Chen, Kudugunta,
  Arivazhagan, and Wu}]{Siddhant2020LeveragingMD}
Aditya Siddhant, Ankur Bapna, Yuan Cao, Orhan Firat, Mia Chen, Sneha Kudugunta,
  Naveen Arivazhagan, and Yonghui Wu. 2020.
\newblock \href {https://doi.org/10.18653/v1/2020.acl-main.252} {Leveraging
  monolingual data with self-supervision for multilingual neural machine
  translation}.
\newblock In \emph{Proceedings of the 58th Annual Meeting of the Association
  for Computational Linguistics}, pages 2827--2835, Online. Association for
  Computational Linguistics.

\bibitem[{Srinivasan et~al.(2021)Srinivasan, Raman, Chen, Bendersky, and
  Najork}]{srinivasan2021wit}
Krishna Srinivasan, Karthik Raman, Jiecao Chen, Michael Bendersky, and Marc
  Najork. 2021.
\newblock \href {https://arxiv.org/abs/2103.01913} {{WIT}: Wikipedia-based
  image text dataset for multimodal multilingual machine learning}.
\newblock \emph{arXiv preprint arXiv:2103.01913}.

\bibitem[{Srivastava et~al.(2020)Srivastava, Mukhopadhyay, K~R, and
  Jawahar}]{Srivastava2020IndicSpeechTC}
Nimisha Srivastava, Rudrabha Mukhopadhyay, Prajwal K~R, and C~V Jawahar. 2020.
\newblock \href {https://aclanthology.org/2020.lrec-1.789} {{I}ndic{S}peech:
  Text-to-speech corpus for {I}ndian languages}.
\newblock In \emph{Proceedings of the 12th Language Resources and Evaluation
  Conference}, pages 6417--6422, Marseille, France. European Language Resources
  Association.

\bibitem[{Tan and Le(2019)}]{Tan2019EfficientNetRM}
Mingxing Tan and Quoc~V. Le. 2019.
\newblock \href {http://proceedings.mlr.press/v97/tan19a.html} {Efficientnet:
  Rethinking model scaling for convolutional neural networks}.
\newblock In \emph{Proceedings of the 36th International Conference on Machine
  Learning, {ICML} 2019, 9-15 June 2019, Long Beach, California, {USA}},
  volume~97 of \emph{Proceedings of Machine Learning Research}, pages
  6105--6114. {PMLR}.

\bibitem[{Tiedemann(2012)}]{TIEDEMANN12.463}
J{\"o}rg Tiedemann. 2012.
\newblock \href
  {http://www.lrec-conf.org/proceedings/lrec2012/pdf/463_Paper.pdf} {Parallel
  data, tools and interfaces in {OPUS}}.
\newblock In \emph{Proceedings of the Eighth International Conference on
  Language Resources and Evaluation ({LREC}'12)}, pages 2214--2218, Istanbul,
  Turkey. European Language Resources Association (ELRA).

\bibitem[{Tyers and Alperen(2010)}]{tyers2010south}
Francis~M Tyers and Murat~Serdar Alperen. 2010.
\newblock South-{E}ast {E}uropean {T}imes: A parallel corpus of {B}alkan
  languages.
\newblock In \emph{Proceedings of the LREC workshop on exploitation of
  multilingual resources and tools for Central and (South-) Eastern European
  Languages}, pages 49--53.

\bibitem[{Yang et~al.(2019{\natexlab{a}})Yang, {\'{A}}brego, Yuan, Guo, Shen,
  Cer, Sung, Strope, and Kurzweil}]{yang-ijcai2019-746}
Yinfei Yang, Gustavo~Hern{\'{a}}ndez {\'{A}}brego, Steve Yuan, Mandy Guo,
  Qinlan Shen, Daniel Cer, Yun{-}Hsuan Sung, Brian Strope, and Ray Kurzweil.
  2019{\natexlab{a}}.
\newblock \href {https://doi.org/10.24963/ijcai.2019/746} {Improving
  multilingual sentence embedding using bi-directional dual encoder with
  additive margin softmax}.
\newblock In \emph{Proceedings of the Twenty-Eighth International Joint
  Conference on Artificial Intelligence, {IJCAI} 2019, Macao, China, August
  10-16, 2019}, pages 5370--5378. ijcai.org.

\bibitem[{Yang et~al.(2020)Yang, Cer, Ahmad, Guo, Law, Constant,
  Hernandez~Abrego, Yuan, Tar, Sung, Strope, and
  Kurzweil}]{yang2019multilingual}
Yinfei Yang, Daniel Cer, Amin Ahmad, Mandy Guo, Jax Law, Noah Constant, Gustavo
  Hernandez~Abrego, Steve Yuan, Chris Tar, Yun-hsuan Sung, Brian Strope, and
  Ray Kurzweil. 2020.
\newblock \href {https://doi.org/10.18653/v1/2020.acl-demos.12} {Multilingual
  universal sentence encoder for semantic retrieval}.
\newblock In \emph{Proceedings of the 58th Annual Meeting of the Association
  for Computational Linguistics: System Demonstrations}, pages 87--94, Online.
  Association for Computational Linguistics.

\bibitem[{Yang et~al.(2019{\natexlab{b}})Yang, Zhang, Tar, and
  Baldridge}]{yang-etal-2019-paws}
Yinfei Yang, Yuan Zhang, Chris Tar, and Jason Baldridge. 2019{\natexlab{b}}.
\newblock \href {https://doi.org/10.18653/v1/D19-1382} {{PAWS}-{X}: A
  cross-lingual adversarial dataset for paraphrase identification}.
\newblock In \emph{Proceedings of the 2019 Conference on Empirical Methods in
  Natural Language Processing and the 9th International Joint Conference on
  Natural Language Processing (EMNLP-IJCNLP)}, pages 3687--3692, Hong Kong,
  China. Association for Computational Linguistics.

\bibitem[{Yoshikawa et~al.(2017)Yoshikawa, Shigeto, and
  Takeuchi}]{yoshikawa2017stair}
Yuya Yoshikawa, Yutaro Shigeto, and Akikazu Takeuchi. 2017.
\newblock \href {https://doi.org/10.18653/v1/P17-2066} {{STAIR} captions:
  Constructing a large-scale {J}apanese image caption dataset}.
\newblock In \emph{Proceedings of the 55th Annual Meeting of the Association
  for Computational Linguistics (Volume 2: Short Papers)}, pages 417--421,
  Vancouver, Canada. Association for Computational Linguistics.

\bibitem[{You et~al.(2020)You, Li, Reddi, Hseu, Kumar, Bhojanapalli, Song,
  Demmel, Keutzer, and Hsieh}]{You2020LargeBO}
Yang You, Jing Li, Sashank~J. Reddi, Jonathan Hseu, Sanjiv Kumar, Srinadh
  Bhojanapalli, Xiaodan Song, James Demmel, Kurt Keutzer, and Cho{-}Jui Hsieh.
  2020.
\newblock \href {https://openreview.net/forum?id=Syx4wnEtvH} {Large batch
  optimization for deep learning: Training {BERT} in 76 minutes}.
\newblock In \emph{8th International Conference on Learning Representations,
  {ICLR} 2020, Addis Ababa, Ethiopia, April 26-30, 2020}. OpenReview.net.

\bibitem[{Young et~al.(2014)Young, Lai, Hodosh, and
  Hockenmaier}]{young2014image}
Peter Young, Alice Lai, Micah Hodosh, and Julia Hockenmaier. 2014.
\newblock \href {https://doi.org/10.1162/tacl_a_00166} {From image descriptions
  to visual denotations: New similarity metrics for semantic inference over
  event descriptions}.
\newblock \emph{Transactions of the Association for Computational Linguistics},
  2:67--78.

\bibitem[{Zhang et~al.(2020)Zhang, Williams, Titov, and
  Sennrich}]{zhang-etal-2020-improving}
Biao Zhang, Philip Williams, Ivan Titov, and Rico Sennrich. 2020.
\newblock \href {https://doi.org/10.18653/v1/2020.acl-main.148} {Improving
  massively multilingual neural machine translation and zero-shot translation}.
\newblock In \emph{Proceedings of the 58th Annual Meeting of the Association
  for Computational Linguistics}, pages 1628--1639, Online. Association for
  Computational Linguistics.

\bibitem[{Zhou et~al.(2021)Zhou, Zhou, Wang, Cheng, Li, Yu, and
  Liu}]{zhou2021uc2}
Mingyang Zhou, Luowei Zhou, Shuohang Wang, Yu~Cheng, Linjie Li, Zhou Yu, and
  Jingjing Liu. 2021.
\newblock Uc2: Universal cross-lingual cross-modal vision-and-language
  pre-training.
\newblock In \emph{Proceedings of the IEEE/CVF Conference on Computer Vision
  and Pattern Recognition}, pages 4155--4165.

\end{thebibliography}
\bibliographystyle{acl_natbib}
\clearpage
\newpage
\setcounter{page}{1}
\appendix

\section{Supplementary Material}

\subsection{Modeling}
\label{appendix:modeling}

\paragraph{Model variants} We include further details about the main model variants we explore:

\textbf{\alignmling}: We use EfficientNet-B5 for the image encoder and BERT-Base Transformer for the text encoder which uses 12 layers, 12 attention heads resulting in an embedding of 768-dimensions. To match the image representation dimension of 512, we add an additional FC layer on top of the text encoder. The \alignmling model has 300M parameters in total, including 30M for EfficientNet-B5, 192M for the token embeddings, and 78M for the BERT Transformer. With this setting, we train on both the full multilingual Alt-Text dataset and the English subset, to get \alignmling and \alignen, respectively. 

\textbf{\muralbase}: The same as \alignmling, but also using text-text learning and the additional projection head for the image-text task (an FC layer that projects the text embedding from 768d to 512d). \textbf{\murallarge}: We use EfficientNet-B7 for the image encoder and BERT-Large Transformer\footnote{24 layers, 24 attention heads, and 1024 hidden size.} for the text encoder. To fit this model into memory, we use a 256-dimension token embedding size and project it to 1024 hidden size, which is then used by the large transformer encoder. The model uses 66M parameters for EfficientNet-B7, 64M for the token embeddings, and 300M for the BERT Transformer (=430M parameters total).

\textbf{\alignhuge} uses an EfficientNet-L2 (=480M parameters) image encoder with a BERT-Large Transformer (300M parameters) as a text encoder. Along with the 64M parameters for token embeddings, \alignhuge has 840M parameters.

\begin{figure}[h!]
    \begin{center}
        \centering
        \includegraphics[scale=0.52]{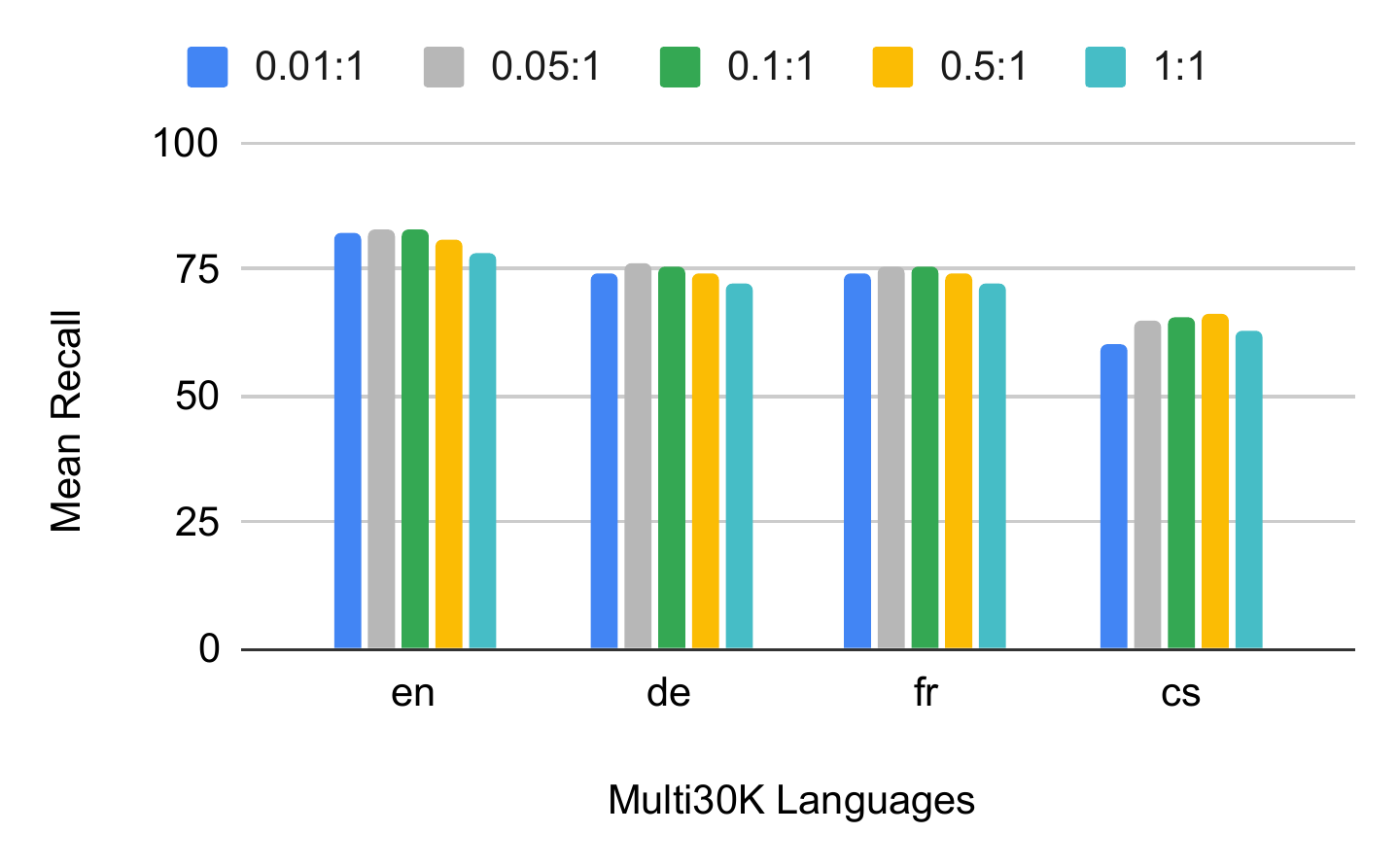}
        \caption{Zero-shot performance on Multi30K (val set) for different task weights (format: text-text weight {:} image-text weight). Overall, a ratio of 0.1:1 works best across all languages.}
    \label{fig:task_weights}
    \vspace{-0.2in}
    \end{center}
\end{figure}

\paragraph{Projection Heads}
For MURAL, we experiment with different layers of projection heads, e.g. 1 Fully Connected (FC) layer and a Multi-Layer Perceptron with non-linearity in between the FC layers. Empirically, we find that MURAL learns better image-text representations when using single layer projection heads on top of the text-encoder, one per task.

\paragraph{Different Task Weights}
Figure \ref{fig:task_weights} shows retrieval performance of models trained using different task weights in the loss function. We report zero-shot results on Multi30K val set for comparison.  Weighing both t2i and i2t tasks equally (1:1) shows a consistent drop in cross-modal retrieval performance, which indicates that we need to weigh text-image task higher than the text-text task for optimal performance. From the figure we see that the ratios 0.1:1 and 0.05:1 achieve similar mean recall for t2t and i2t tasks across all Multi30K languages. In all our experiments, we use the ratio 0.1:1 for training MURAL.

\paragraph{Checkpoint Initialization.} For MURAL, we either (1) initialize from a trained ALIGN checkpoint or (2) train both encoders from scratch. Our early experiments showed that the first strategy does not work as well. This is likely because ALIGN discards information about other languages early on because of English dominance in the Alt-Text dataset (\ref{fig:wit_alttxt_stats})--and as a result, performance on other languages is worse when training with a multitask objective. Since the model training with checkpoint initialization achieves a higher performance faster than the model trained on scratch, it offers a potential trade-off between performance and time for training. Given the early empirical results, in this paper, we always train MURAL from scratch unless otherwise stated. We stress that in the MURAL multitask model, the per-task layers on top of the text-encoders are trained from scratch in both the settings.

\paragraph{Finetuning Strategies: Single-task vs. Multi-task}
We experimented with the standard single-task fine-tuning using image-text pairs in downstream datasets like Multi30K. However, we also tried constructing text-text aligned pairs from the Multi30K dataset (e.g. by using co-caption pairs as text-text pairs), similar to the multitask strategy of \citet{parekh-etal-2021-crisscrossed}. We found that including text-text fine-tuning slightly decreased cross-modal retrieval performance. This is may be because the large pretrained MURAL model benefits little from seeing text-text pairs at the fine-tuning stage. This is interesting because this indicates that the training strategies at different stages affect the final performance differently. That said, it may just be that we lack the necessary evaluation data, such as multilingual variant of Crisscrossed Captions \cite{parekh-etal-2021-crisscrossed} with non-English Semantic Textual Similarity scores. 

\begin{figure*}[t]
    \centering
    \begin{subfigure}[b]{0.49\textwidth}
         \centering
         \includegraphics[width=\textwidth]{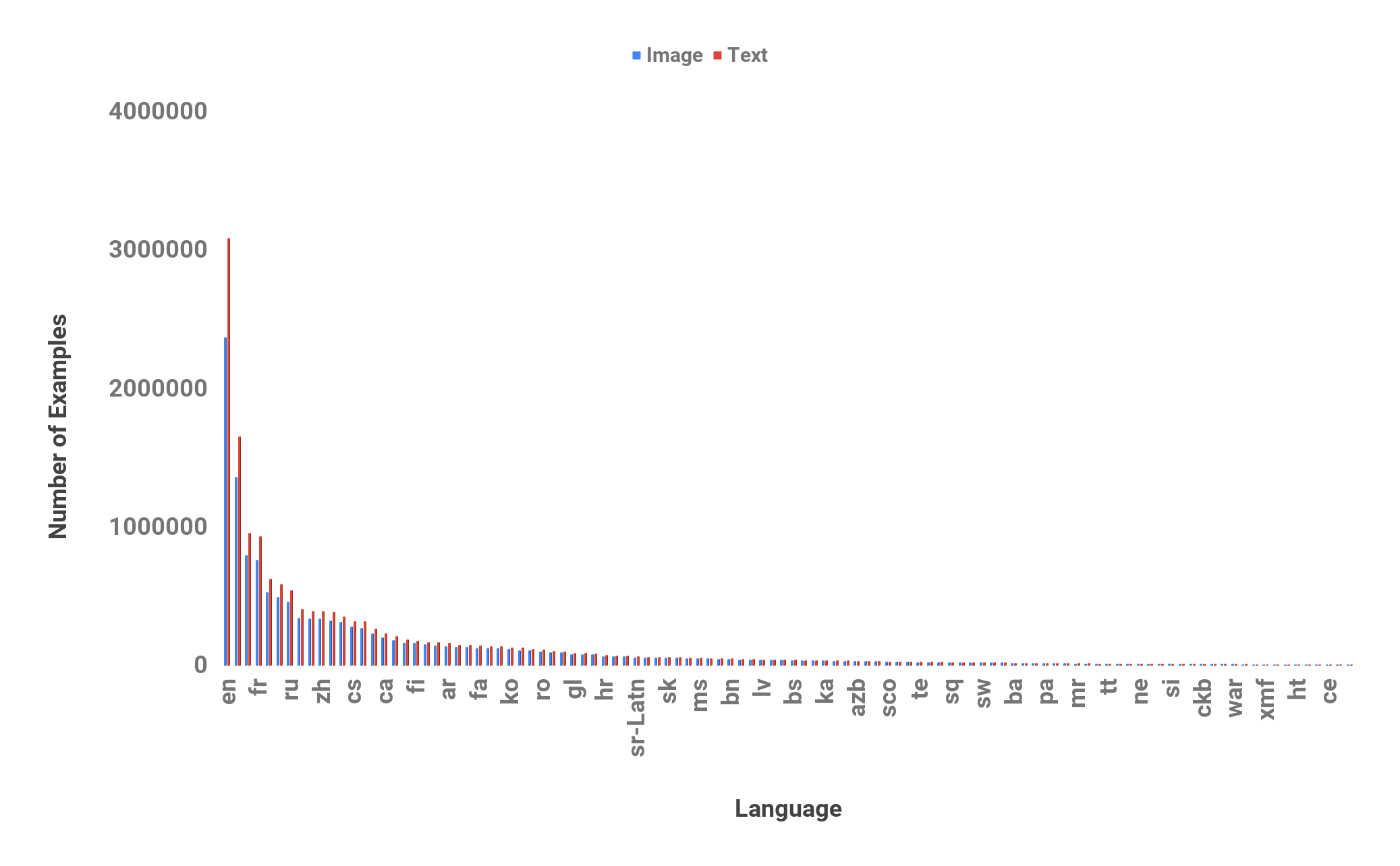}
        %  \caption{Log Scale}
         \label{fig:y equals x}
     \end{subfigure}
    % \hfill
    \begin{subfigure}[b]{0.49\textwidth}
         \centering
         \includegraphics[width=\textwidth]{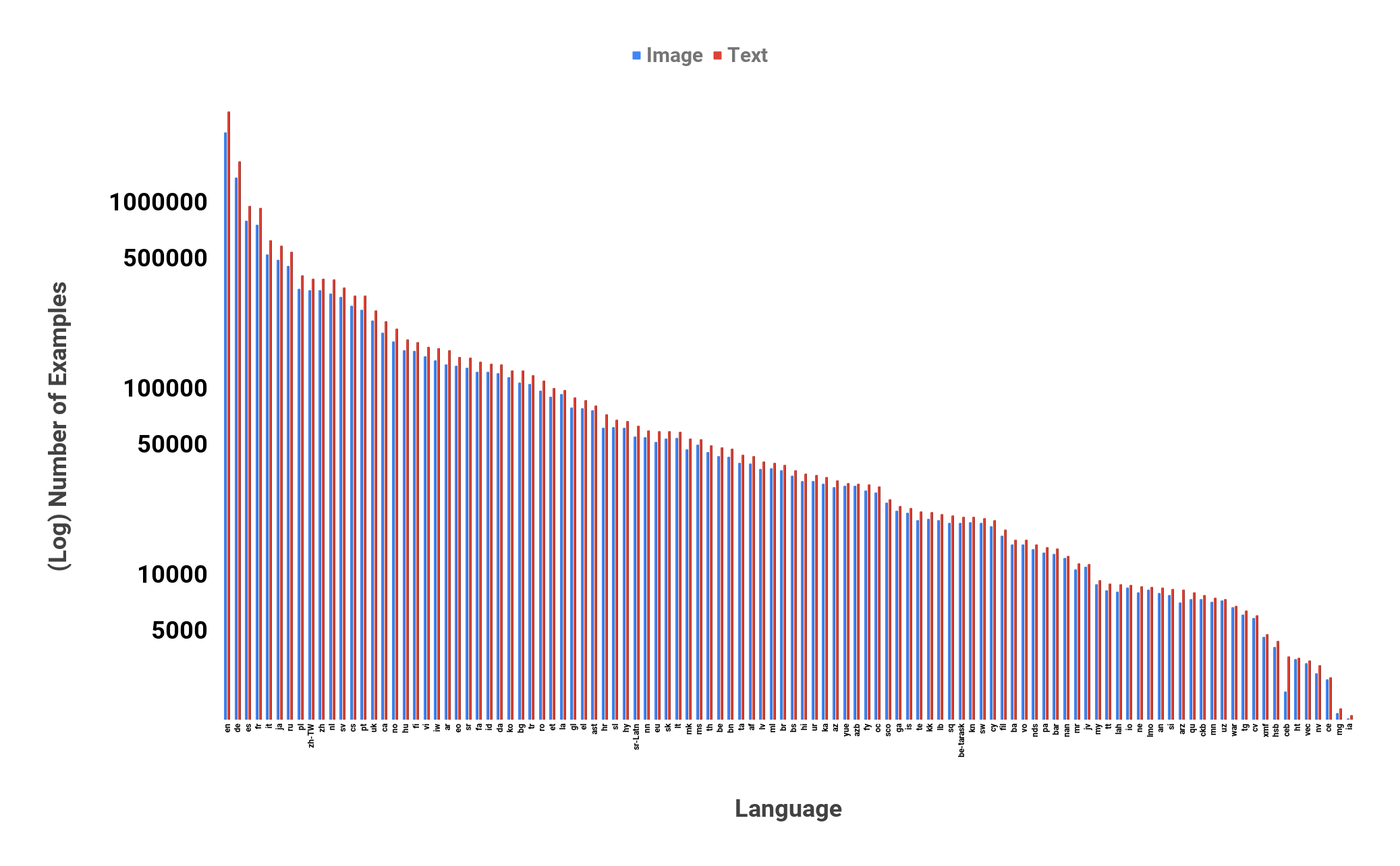}
        %  \caption{Linear Scale}
         \label{fig:y equals x}
    \end{subfigure}
     
    \caption{WIT language distribution: (left) linear scale, which clearly conveys the skew toward well-resourced languages; (right) log-scale, which provides a better view of under-represented languages.}
    \label{fig:wit_stats}
\end{figure*}

\subsection{Ensemble of Open  Bilingual Translation  (EOBT)  Pairs}
\label{appendix:eobt}
The complete list of open-sourced bilingual translation pairs dataset used in the construction of EOBT includes: Europarl \cite{koehn2005europarl}, Paracrawl \cite{espla-etal-2019-paracrawl}, TED57, Tanzil \cite{TIEDEMANN12.463}, NewsCommentary, Wikimatrix \cite{schwenk2019wikimatrix}, Wikititles, JW300 \cite{agic2020jw300}, Opus100 \cite{zhang-etal-2020-improving}, SETimes \cite{tyers2010south}, UNv1.0, Autshumato \cite{groenewald2010processing}, PMIndia \cite{haddow2020pmindia}, CVIT \cite{Srivastava2020IndicSpeechTC}, Inuktitut \cite{hernandez2020ubiqus}, NLPC, JESC \cite{pryzant_jesc_2018}, KFTT \cite{neubig11kftt}, ASPEC \cite{Nakazawa2016ASPECAS}, Flores \cite{flores_dataset}. The data was processed in the same way as outlined in \citet{Siddhant2020LeveragingMD}.

\subsection{Wikipedia Image-text Dataset}
\label{appendix:wit}
To maintain high quality text descriptions, all the splits in the WIT dataset uses the reference descriptions paired with the images. This is the text description underneath an image in a Wikipedia page. This also prevents any potential overlap with the Alt-Text training data. Similar to the Alt-Text data distribution across languages, WIT data distribution (\ref{fig:wit_stats}) is heavily skewed in favor of well-resourced languages. Refer to the \citet{srinivasan2021wit} for more details on dataset collection and statistics. Since WIT's test set has been withheld for a competition, we use only the publicly available training set of approximately ~37M image-text examples with ~11M images. The actual available data is reduced because of our use of only reference description text as there are only about ~16M reference descriptions in the WIT dataset. We split this into 108 individual language sets based on the language of the Wikipedia page. We observe that sometimes a particular language page might include a caption in an alternate language, especially an under-resourced language using a text in an well-resourced language. For e.g., an image in a Hindi page has a text caption in English. Each language set is further split into train, val and test sets. We maintain 5K image-text pairs for most of the languages but for the under-resourced we cut this down to 3K or 1K. For each language, we make sure that an image is only in one set (train, val, test).

We also create two evaluation groups from WIT for well-resourced languages and under-resourced ones, ensuring they cover a broad range of language families and geographic areas:

\begin{itemize}
    \item \textbf{well-resourced}: English (en), German(de), French (fr), Czech (cs), Japanese (ja), Chinese (zh), Russian (ru), Polish (pl), Turkish (tr)
    \item \textbf{under-resourced}: Tajik (tg), Uzbek (uz),  Irish (ga), Belarusian (be), Malagasy (mg), Cebuano (ceb),  Haitian (ht), Waray-Waray (war)
\end{itemize}

% \begin{figure*}
%     \begin{center}
%         \includegraphics[width=\textwidth]{images/wit_log.png}
%         \caption{WIT Dataset Image-Text distribution per language plotted using log-scale.}
%     \label{fig:wit_stats}
%     \vspace{-0.2in}
%     \end{center}
% \end{figure*}

% \begin{figure*}
%     \begin{center}
%         \includegraphics[width=\textwidth]{images/alt_log2.png}
%         \caption{Alt-Text Dataset distribution per language plotted using log-scale.}
%     \label{fig:alttext_stats}
%     \vspace{-0.2in}
%     \end{center}
% \end{figure*}

% \begin{figure*}
%     \begin{center}
%         \includegraphics[width=0.5\textwidth]{images/wit_linear.png}
%         \caption{WIT Dataset Image-Text distribution per language plotted linearly to highlight the skew towards well-resourced languages.}
%     \label{fig:wit_stats_linear}
%     \vspace{-0.2in}
%     \end{center}
% \end{figure*}

% \begin{figure*}
%     \begin{center}
%         \includegraphics[width=\textwidth]{images/alt_linear.png}
%         \caption{Alt-Text Dataset distribution per language plotted linearly to highlight the skew towards well-resourced languages.}
%     \label{fig:alttext_stats_linear}
%     \vspace{-0.2in}
%     \end{center}
% \end{figure*}

\begin{table}
  \caption{Image-Text data size distribution across languages for WIT and Alt-Text Datasets}
  \vspace{-0.5em}
  \label{tab:wit-alt-text-aggr-stats}
  \begin{tabular}{ccc}
%    \hline 
    {\# Examples} 
    & {Alt-Text \# Lang}
    & {WIT \# Lang} \\ 
  \hline
 $> 10^8$ & 4 & - \\
%  \hline
 $> 10^7$ & 11 & - \\
%  \hline
 $> 10^6$ & 22 & 2 \\
%  \hline
 $> 10^5$ & 37 & 29 \\
%  \hline
 $> 10^4$ & 18 & 52 \\
%  \hline
 $> 10^3$ & 12 & 25 \\
%  \hline
 $> 10^2$ &  4 & - \\
%  \hline
 $> 10^1$ &  2 & - \\
  \hline
{Total} & {110} & {108} \\
%  \hline
\end{tabular}
\end{table}

\subsection{Translate-Train Languages}
\label{appendix:translate_train}
For translate-train baseline, we translate the English captions to some other well-resourced languages. For Alt-Text translation we translate English Alt-Text to German, French, Czech, Japanese, Korean, and Chinese. For CC12m dataset, we translate to languages present in the Multi30k and MSCOCO dataset namely, German, French, Czech, and Japanese. We augment the image-text pairs in English with these machine translated captions for training.

\begin{figure}[t]
    \centering
    \includegraphics[width=0.47\textwidth]{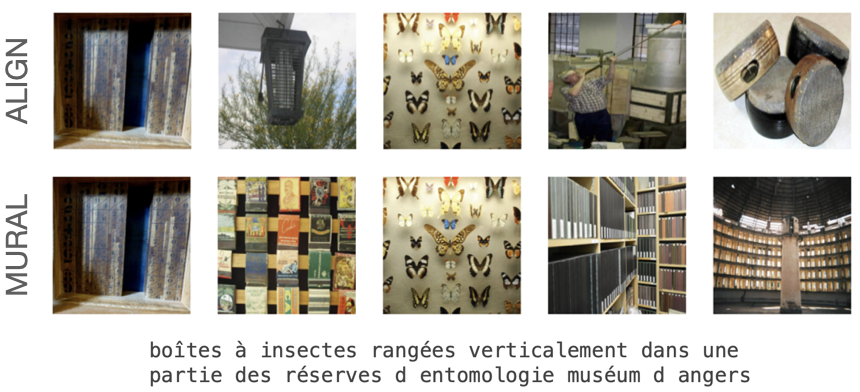}
    \caption{Fidelity to word `boîtes` (boxes) in a French caption}
    \label{fig:appendix_fidelity1}
\end{figure}

\begin{figure}[t]
    \centering
    \includegraphics[width=0.47\textwidth]{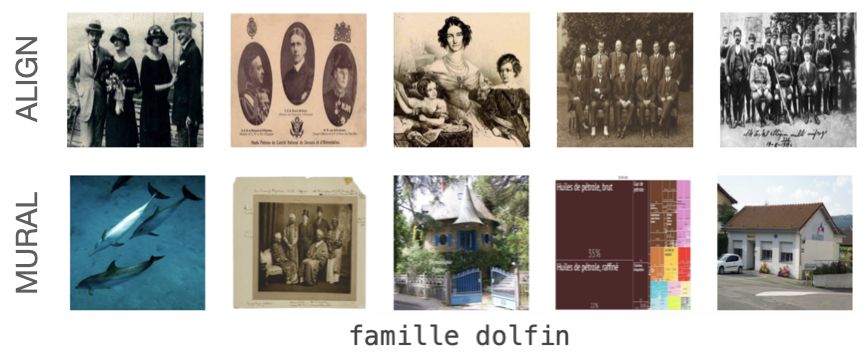}
    \caption{Fidelity to both words famille and dolfin with MURAL}
    \label{fig:appendix_fidelity2}
\end{figure}

\begin{figure}[t]
    \centering
    \includegraphics[width=0.47\textwidth]{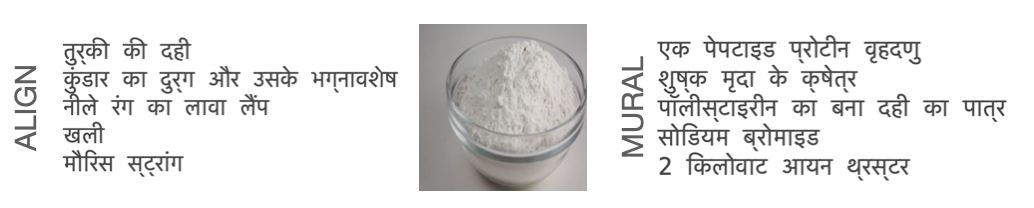}
    \caption{Color identification of the image to retrieve captions describing food that matches the white color represented in the image}
    \label{fig:appendix_color2}
\end{figure}

\begin{figure}[t]
    \centering
    \includegraphics[width=0.47\textwidth]{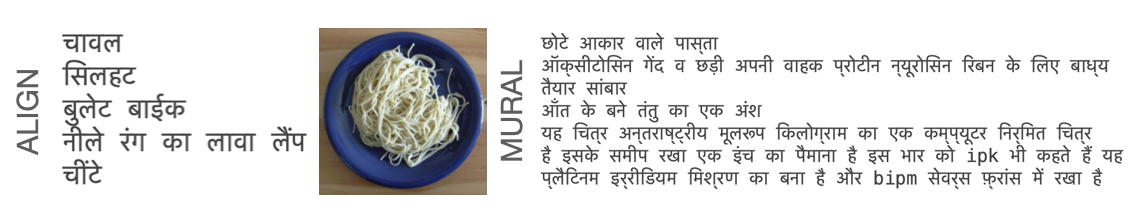}
    \caption{Identifying the noodles by its color and shape to retrieve captions such as "rice".}
    \label{fig:appendix_color3}
\end{figure}

\begin{figure}[h!]
    \centering
    \includegraphics[width=0.47\textwidth]{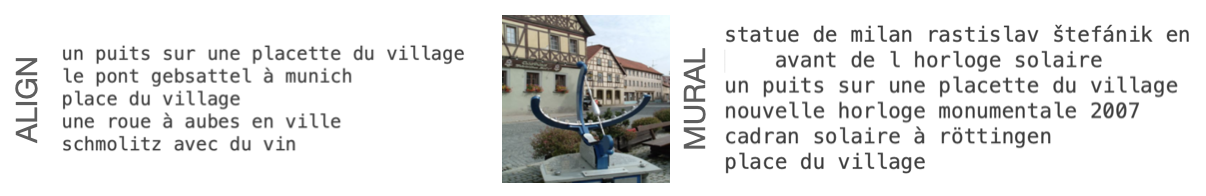}
    \caption{MURAL learns to identify the sundial ("cadran solaire" in French) being displayed in the input image}
    \label{fig:appendix_identfi1}
\end{figure}

\begin{figure}[h!]
    \centering
    \includegraphics[width=0.47\textwidth]{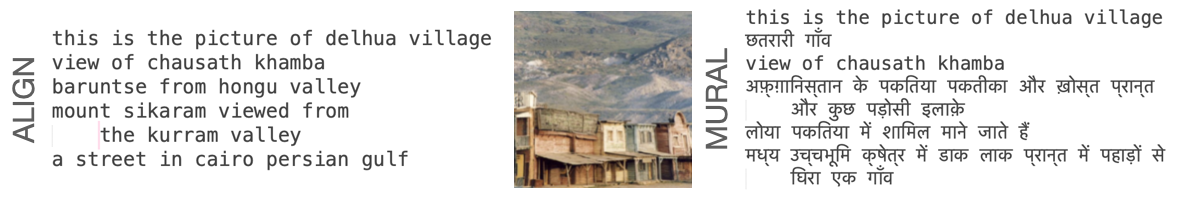}
    \caption{For an input image, both ALIGN and MURAL tend to retrieve English captions than Hindi captions}
    \label{fig:appendix_lang_id}
\end{figure}

\subsection{Error Analysis}
\label{appendix:error_analysis}
We include more examples of retrieved images and text on the WIT dataset comparing ALIGN and MURAL. Some more observations-

Using color as pivots is displayed by both ALIGN and MURAL in retrieving examples, but is stronger in MURAL. For instance (Figure \ref{fig:appendix_color2}), identifying image of flour by its color. Also in Figure \ref{fig:appendix_color3}, ALIGN uses white and blue to retrieve captions mentioning those colors. This kind of backfires for ALIGN, because it retrieves "Blue colored lava lamp" as one of the captions.
With MURAL we observe an increased object identification performance. In Figure \ref{fig:appendix_identfi1}, ALIGN fails to identify the sundial in the image, whereas MURAL retrieves the correct caption. We believe additional translation pairs helped MURAL learn the word for sundial in French.

For a relatively under-resourced language such as Hindi, both ALIGN and MURAL have a tendency to retrieve captions in English, which is comparatively high-resourced (Figure \ref{fig:appendix_lang_id}. However, in comparison to ALIGN, MURAL tends to infer characters and culture from the images and retrieve more Hindi captions.

Some of these observations hint us that there is definite value in using translation data to improve representations for which data is scarce. We see there are clear benefits of MURAL over ALIGN for languages other than English.

\end{document}